\documentclass[sigconf,nonacm]{acmart}
\AtBeginDocument{%
  }
\newcommand{\para}[1]{\noindent\textbf{#1.}}
\usepackage{listings}
\usepackage{array}
\usepackage{graphicx}
\usepackage{multirow}
\usepackage[switch]{lineno}
\usepackage{float}
\usepackage{algorithm}
\usepackage{algpseudocode}
\newcommand{\transql}{TranSQL$^{+}$\,}
\usepackage{enumitem}
\usepackage{balance}

\setcopyright{acmlicensed}
\copyrightyear{2018}
\acmYear{2018}
\acmDOI{XXXXXXX.XXXXXXX}
\acmConference[Conference acronym 'XX]{Make sure to enter the correct
  conference title from your rights confirmation email}{June 03--05,
  2018}{Woodstock, NY}
\acmISBN{978-1-4503-XXXX-X/2018/06}

\begin{document}

\title{TranSQL$^{+}$: Serving Large Language Models with SQL on Low-Resource Hardware
}



\author{Wenbo Sun}
\affiliation{%
  \institution{Delft University of Technology}
  \country{Netherlands}
}
\email{w.sun-2@tudelft.nl}

\author{Qiming Guo}
\affiliation{%
  \institution{Texas A\&M University - Corpus Christi}
  \country{USA}
}
\email{guoqm07@gmail.com}

\author{Wenlu Wang}
\affiliation{%
  \institution{Texas A\&M University - Corpus Christi}
  \country{USA}
}
\email{wenlu.wang@tamucc.edu}

\author{Rihan Hai}
\affiliation{%
  \institution{Delft University of Technology}
  \country{Netherlands}
}
\email{r.hai@tudelft.nl}


\renewcommand{\shortauthors}{Trovato et al.}

\begin{abstract} 
Deploying Large Language Models (LLMs) on resource-constrained devices remains challenging due to limited memory, lack of GPUs, and the complexity of existing runtimes. In this paper, we introduce \textbf{\transql}, a template-based code generator that translates LLM computation graphs into pure SQL queries for execution in relational databases. Without relying on external libraries, \transql\, leverages mature database features—such as vectorized execution and out-of-core processing—for efficient inference. We further propose a row-to-column (ROW2COL) optimization that improves join efficiency in matrix operations. Evaluated on Llama3-8B and DeepSeekMoE models, \transql achieves up to $20\times$ lower prefill latency and $4\times$ higher decoding speed compared to DeepSpeed Inference and \texttt{Llama.cpp} in low-memory and CPU-only configurations. Our results highlight relational databases as a practical environment for LLMs on low-resource hardware. 
\end{abstract}




\maketitle

\section{Introduction}
Large Language Models (LLMs) have demonstrated remarkable capabilities, yet they are predominantly deployed on powerful cloud servers with high-end GPUs and abundant memory resources.  The cloud‑centric paradigm raises privacy concerns and complicates integration with local tools. When a task requires local tooling, one must either expose the local environment \cite{privacy1,privacy2} so a remote service can produce tool‑compatible outputs or accept reduced functionality—undermining both privacy and customizability. These drawbacks motivate performing LLM inference directly on resource‑constrained personal or edge devices. With appropriate fine-tuning or distillation, edge deployments can provide sufficient performance for many specialized applications. For instance, multiple studies in healthcare~\cite{blade,medmoe} have demonstrated that smaller, domain-specific models can achieve superior accuracy compared to large general-purpose models, enabling specialized deployments such as on-device diagnostic tools or personal health assistants.

Nevertheless, edge devices—ranging from personal computers to billions of smartphones and IoT endpoints—typically lack GPUs, have limited memory, and operate within fragmented software and hardware ecosystems. Crucially, even smaller domain-specific models can exceed the memory of typical edge devices. For example, Llama3 8B \cite{llama3}, which is often praised for its performance–size tradeoff, requires 16.1GB of RAM when loaded in memory, exceeding the 8–16GB available on most laptops and smartphones. This constraint complicates efficient single-batch inference under memory limitations, raising a significant challenge for practical edge-based LLM deployments.

A typical LLM deployment on edge devices starts with model compression, followed by compilation to map neural operators to hardware-specific tensor backends. Most setups load the entire model into memory for fast inference, but this is often infeasible on devices with limited RAM. To address this, various techniques have been proposed—each with trade-offs. Compression methods (e.g., distillation, quantization, pruning) reduce memory usage but can degrade accuracy and require tuning~\cite{powerinfer,llmflash}. Offloading weights between GPU and CPU memory~\cite{flexgen,vllm,deepspeed} is another strategy, though mainly optimized for GPU systems. Few frameworks support weight offloading directly to NVMe storage. The frameworks, such as \texttt{Llama.cpp}, rely on memory-mapped I/O to support swapping active weights from disk to memory dynamically~\cite{llamacpp,flexinfer,llmflash,deepspeed}. However, NVMe offloading in these systems often introduces significant I/O bottlenecks and inefficient disk access patterns, severely degrading inference performance when memory is constrained.

In addition, the heterogeneity of edge environments complicates deployment. Personal computers are predominantly x86\_64, whereas embedded and edge devices span ARM v8/v9 \cite{armv9}, RISC-V \cite{riscv}, and other ISAs. Although deep learning compilers (e.g., ONNX \cite{onnx}, TVM \cite{tvm}, MLIR \cite{mlir}) can compile models to common intermediate representations, supporting new operators and heterogeneous hardware still demands re-engineering and related skills: developers must implement ISA-specific kernels and instruction support, then ensure every target device matches up-to-date runtime. Consequently, porting and optimizing inference frameworks across such diverse hardware remains complex and expensive.


\para{Our proposal} To address these challenges, we propose a fundamentally different approach: leveraging the ubiquity and maturity of \textbf{relational databases} to serve as inference engines. Databases such as SQLite are embedded in billions of devices worldwide \cite{lite}, while modern database engines offer robust out-of-core execution, efficient cache management, and vectorized execution~\cite{vectorized1,vectorized2}. Databases align well with the demands of interactive workloads such as LLM inference under tight memory budgets, efficiently managing model parameters that exceed RAM and accelerating linear‑algebra kernels via batch‑oriented, vectorized execution. The same buffer‑management and I/O‑scheduling advantages benefit streaming workloads—for example, object detection with convolutional neural networks (CNNs)—making databases a strong fit for resource‑constrained deployments.

Moreover, SQL’s inherent portability enables deployment across diverse platforms without architecture‑specific tweaks or external ML runtimes. Prior work has shown that relational databases can execute neural computations when cast in relational form \cite{duckbrain,declaritive,dl2sql}, confirming their viability as inference engines.  In practice, this means developers with basic database skills can serve models using familiar SQL operations, rather than learning or maintaining specialized compiler stacks.

Building on these observations, we introduce \transql, a template-based code generator that converts LLM computation graphs into pure SQL queries executable entirely within a relational database. The forward pass of the model—including matrix multiplications, attention, and non-linear activations—is expressed using relational operations such as joins and aggregations. This design requires \emph{no external libraries or custom inference engine}, relying entirely on the database engine. Unlike prior in-database ML approaches that depend on external deep learning backend~\cite{sqlserver,postgreml,verticaML}, \transql uses portable SQL, enabling deployment on any standard database engines (e.g., PostgreSQL, DuckDB, Clickhouse) with minimal engineering effort. The method in this paper specifically targets CPU-only, memory-constrained hardware, where LLM inference typically serves a single request at a time, distinguishing it clearly from GPU-accelerated cloud-based inference services. Consequently, this paper's evaluations and optimizations are tailored explicitly to this scenario and are not directly comparable with GPU-oriented deep learning frameworks.

\begin{figure}[!t]
    \centering
    \includegraphics[width=\linewidth]{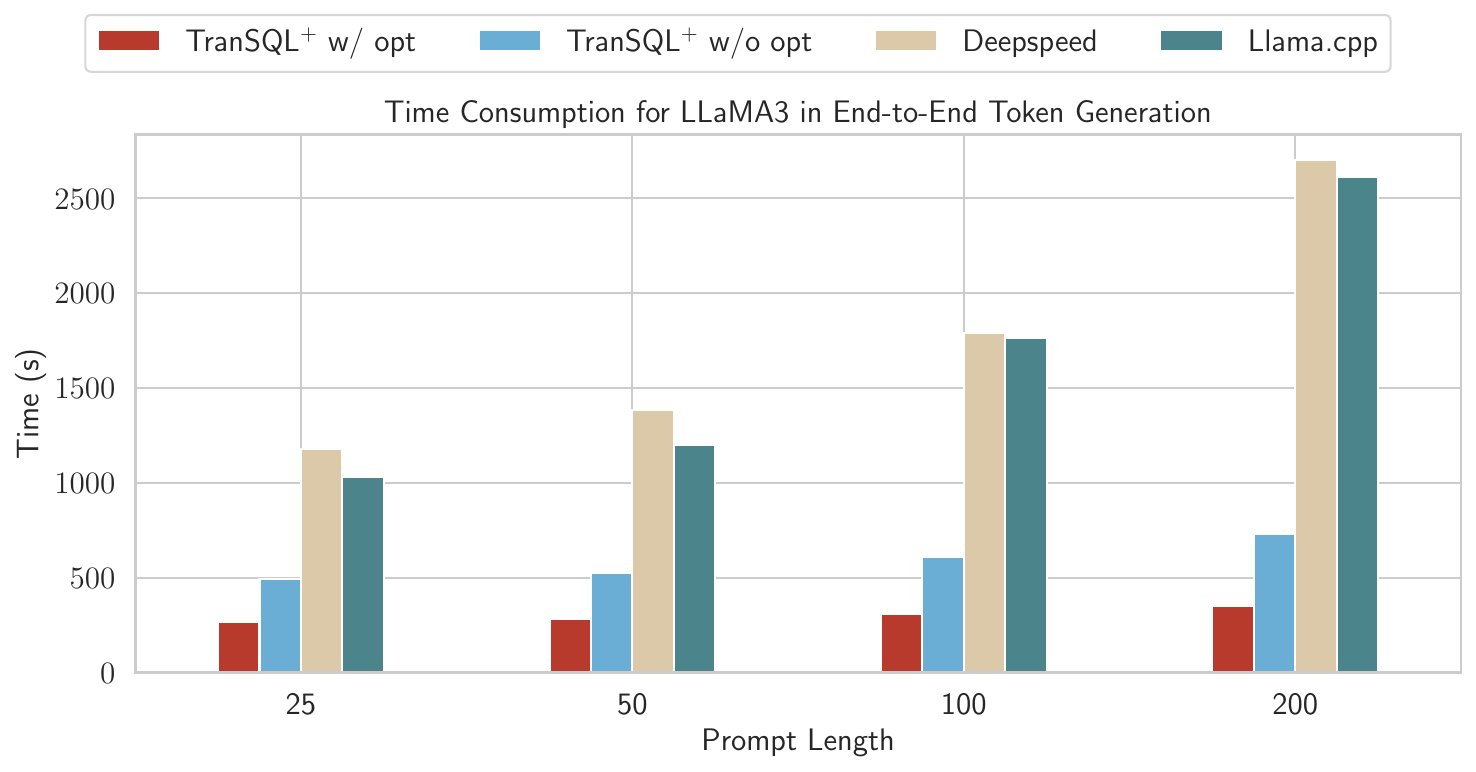}
    \caption{End-to-end time consumption. The output token length is 50. \transql shows \textbf{\boldmath 10$\times$} faster end-to-end (Prefill \& Decoding)  performance than production-ready frameworks.}    
    \label{fig:time_intro}
    \vspace{-3mm}
\end{figure}

\para{Contributions} This paper makes the following key contributions:
\begin{itemize}[leftmargin=*, labelsep=0.5em, itemsep=0.2em]
    \item We present \textit{TranSQL$^{+}$}, the first template-based SQL code generator capable of serving modern LLMs entirely within a relational database engine, requiring \textbf{no external dependencies}.
    \item We introduce \textsc{ROW2COL} and table fusion optimizations that leverage structural insights from matrix multiplication and the execution patterns of vectorized database engines to reduce intermediate result sizes and enhance parallelism. With these optimizations, \transql achieves a 2$\times$ speedup over the unoptimized SQL baseline.
    \item We evaluate \transql on two representative models—\textit{Llama3 8B} and \textit{DeepSeek MoE}—under a CPU-only setting (4 cores, 16GB RAM), showing up to \textbf{10$\times$} speedups in end-to-end token generation compared to production-ready frameworks such as DeepSpeed Inference \cite{deepspeed} and \texttt{Llama.cpp} \cite{llamacpp} (see Fig.~\ref{fig:time_intro}).
\end{itemize}

Our evaluation confirms that relational databases offer a practical, efficient, and highly portable backend for serving LLM inference in memory-constrained CPU-only environments overlooked by traditional frameworks.  

This paper is organized as follows: Sec.~\ref{sec:background} reviews background on LLM inference; Sec.~\ref{sec:transql} describes the design of \transql; Sec.~\ref{sec:opt} details optimization techniques; Sec.~\ref{sec:experiments} presents our experimental results; Sec.~\ref{sec:related} discusses related work; and Sec.~\ref{sec:conclusion} concludes.

\section{Background}
\label{sec:background}
LLMs are built on Transformer \cite{attention} architectures. During inference, the model receives a sequence of input tokens (e.g., words or subwords), encodes them into embeddings, and processes them through multiple layers to iteratively update these representations. The model then generates output tokens one at a time recursively.

This section introduces the fundamental computations involved in LLM inference, highlighting the core neural operators that underpin these models. Understanding these operators provides the foundation for our template-based approach to translating LLM inference into executable SQL queries. In addition to these micro-level operators, we also briefly present two representative model architectures—dense and mixture-of-experts (MoE)—to illustrate how structural differences can lead to varying performance characteristics when serving LLMs within relational databases.

\subsection{LLM Inference Computation}
Large language models (LLMs) like Llama3 \cite{llama3} are implemented as deep transformer networks. Each Transformer layer applies two key sublayers – a self-attention mechanism \cite{attention}  and a position-wise feed-forward network – each with a residual addition.

\para{Self-Attention and GQA} The self-attention mechanism enables each token to attend to all others in a sequence. In a standard \textit{multi-head attention} setup, the model computes separate \textit{query} ($Q$), \textit{key} ($K$), and \textit{value} ($V$) projections for each attention head. Given an input token representation $h$, a single head computes:
\[
Q = h W_Q,\quad K = h W_K,\quad V = h W_V,
\]
where $W_Q$, $W_K$, and $W_V$ are learned projection matrices. Given matrices $Q$, $K$, and $V$ for all tokens, the attention output is computed as:
\begin{equation}
\label{eq:attn}
    \text{Attn}(Q, K, V) = \operatorname{softmax}\left( \frac{Q K^T}{\sqrt{d_k}} \right) V,
\end{equation}
where $d_k$ is the dimensionality of each attention head.

To improve efficiency, recent LLMs such as Llama3 adopt a variant called \textit{Grouped-Query Attention (GQA)} \cite{GQA}. GQA partitions the $H$ attention heads into $G$ groups ($G < H$), where all heads in the same group share their key and value projections. For example, Llama1 \cite{llama1} uses standard multi-head attention, but switches to GQA in the later versions \cite{llama3} to reduce memory bandwidth consumption.

Formally, given queries $Q = [Q^{(1)}, \dots, Q^{(H)}]$, GQA constructs only $G$ distinct key-value pairs $K^{(1)}, \dots, K^{(G)}$ and $V^{(1)}, \dots, V^{(G)}$. Each head $i$ is assigned to a group $g = f(i)$, and performs attention as:
\[
\text{Attn}_i = \operatorname{softmax}\left( \frac{Q^{(i)} (K^{(g)})^T}{\sqrt{d_k}} \right) V^{(g)}.
\]
When $G = H$, this reduces to standard multi-head attention; when $G = 1$, it becomes  \emph{multi-query attention (MQA)} \cite{mqa}—using shared $K, V$ projections for all heads.

We adopt GQA in this work as it represents a more general and efficient formulation. Importantly, both GQA and standard multi-head attention rely on the same underlying neural operators (e.g., matrix multiplication, softmax), making them compatible with our SQL-based code generation framework.

\para{SwiGLU Feed-Forward Activation}
Following attention, each token passes through a position-wise feed-forward network (FFN).  Llama3 uses the \textbf{SwiGLU} \cite{swiglu} activation—a gated, smooth nonlinearity formed by multiplying two linear projections (one passed through Swish). Unlike \textbf{ReLU}, which zeroes all negative inputs, and \textbf{GELU}, which applies a fixed Gaussian-shaped smooth gate, SwiGLU’s learned gating improves gradient flow and expressiveness, yielding consistently better perplexity and convergence in large transformers. The SwiGLU activation is defined as:
\begin{equation}
    \text{SwiGLU}(h) = (hW_1 + b_1) \odot \text{Swish}(hW_2 + b_2),
\end{equation}
where $\odot$ is element-wise multiplication and $\text{Swish}(x) = x \cdot \sigma(x)$ with $\sigma$ being the sigmoid function. The FFN output is then linearly projected back to the model dimension and added to the residual stream.

\smallskip

\para{Summary} Each Llama Transformer layer consists of: (i) input normalization, (ii) multi-head or grouped-query attention, and (iii) a SwiGLU-activated feed-forward network, each has a residual connection to updated input embeddings. These operations form the core of modern LLM inference.
\subsection{Prefill and Decoding}

Token generation in LLM inference consists of two distinct phases: \textbf{prefill} and \textbf{decoding} \cite{prefilldecoding}. During the prefill phase, the model processes the input context to compute the query ($q$), key ($k$), and value ($v$) vectors for all input tokens. This phase is the most resource-intensive, both in terms of computation and memory.

Once the first token is generated, the model enters the decoding phase, where new tokens are generated one at a time. In this phase, the attention vectors of previously generated tokens can be cached. As a result, only the $q$, $k$, and $v$ vectors of the most recently generated token need to be computed, significantly reducing the computation cost compared to the prefill phase.

\subsection{Mixture-of-Experts vs. Dense Transformers}
\begin{figure}[!t]
    \centering
    \includegraphics[width=0.85\linewidth]{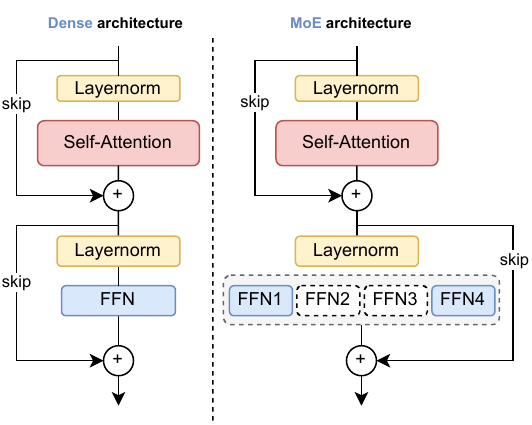}
\caption{Differences of dense models VS. MoE models}    
\label{fig:moe_vs_dense}
\vspace{-3mm}
\end{figure}
An alternative approach to scale LLMs is the \textbf{Mixture-of-Experts (MoE)} architecture \cite{moe}, exemplified by the recent LLMs, such as DeepSeek MoE \cite{deepseekmoe} and QWen2MoE \cite{qwen2}. In a \emph{dense} Transformer (e.g., Llama or GPT-4), every token activates the same set of parameters—typically a full feed-forward network—regardless of the input. In contrast, an \emph{MoE} Transformer consists of multiple parallel feed-forward networks (experts), but only a subset of these experts is activated for any given token (see Fig.~\ref{fig:moe_vs_dense}).

Because only a small fraction of the parameters is active during each forward pass, only the relevant model weights need to be loaded into memory during inference. This makes MoE a compelling choice for deployment in resource-constrained environments, as MoE-based models can achieve similar task performance while running faster than their dense counterparts.

Moreover, MoE architectures are especially well-suited for database-based LLM serving during decoding. By partitioning weights and indexing by expert, databases can selectively load only the needed segments, leveraging native caching and query optimization to reduce I/O and manage memory efficiently. This synergy enables scalable, low-overhead inference. This collaboration between the MoE architecture and the inherent strengths of databases potentially facilitates fast, scalable, and memory-efficient inference.

\section{\transql}
\label{sec:transql}

\begin{figure*}[t]
    \centering
    \includegraphics[width=0.8\linewidth]{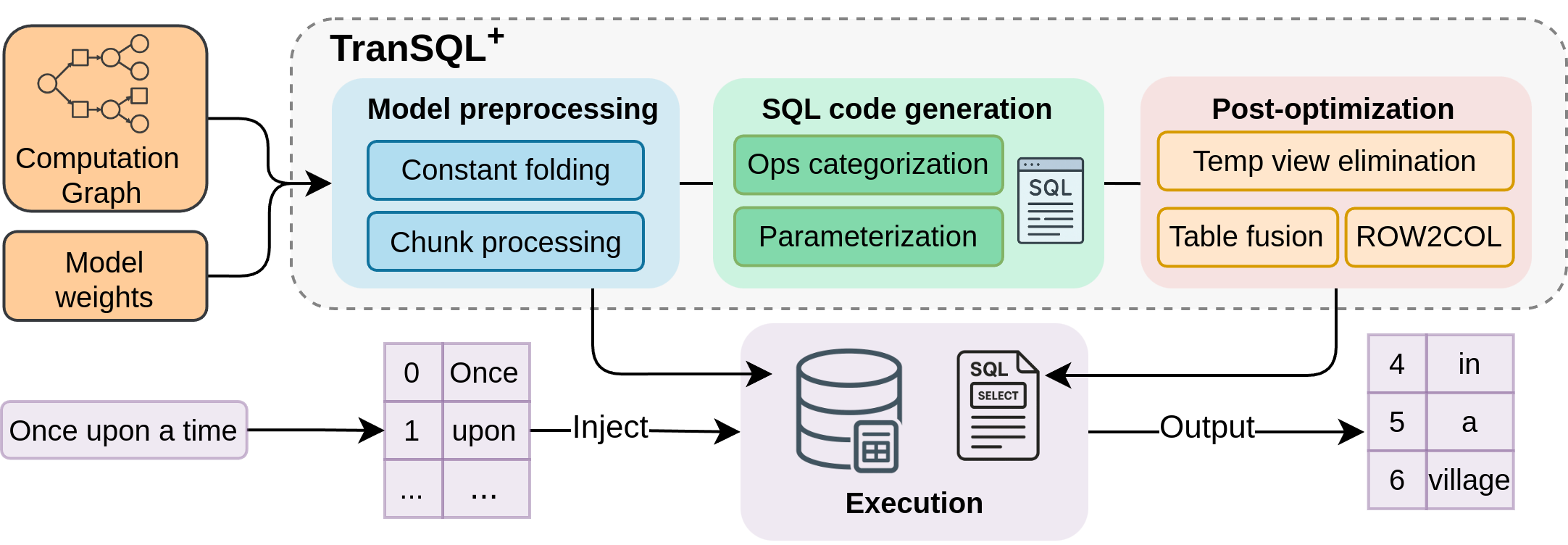}
\caption{An overview of the workflow in \transql.}    
\label{fig:workflow}
\end{figure*}

\transql consists of three main components: \emph{model preprocessing}, \emph{template-based SQL code generation}, and \emph{post-optimization} (see Fig.~\ref{fig:workflow}).  To prepare model data for \transql, we first download the model’s weights and configuration (e.g., embedding size, layer count) from repositories like HuggingFace\footnote{\url{https://huggingface.co/}}.

\para{Model Preprocessing} The forward-pass computation graph is exported via ONNX \cite{onnx} —an open, framework-agnostic neural network interchange format— and simplified using constant folding. Meanwhile, weights are loaded with PyTorch, chunked, mapped to a relational schema, and stored in columnar files (e.g., Parquet). If the input model follows a MoE architecture, \transql additionally builds indexes on expert identifiers to efficiently retrieve only the activated expert weights during inference.  The database files are larger than the raw model tensor files because a database stores additional metadata (schemas, catalogs, pages, indexes, free-space maps, etc.) rather than just contiguous value arrays. Concretely, Llama3 8B grows from 16.1GB (raw) to 21.3GB in the database, and DeepSeek MoE from 32.8GB to 48.6GB, including 0.7GB of indexes for experts. Llama3 uses no index: as a dense model, nearly all weights participate in computation, yielding low selectivity and little benefit from indexing. The preprocessing runs once on a server-class machine when integrating a new model; the resulting SQL queries and data can then be transferred to target devices. 

\para{SQL Code Generation} Given the computation graph of an LLM inference pass, \transql\, extracts all neural operators and maps each to a corresponding SQL template. The generator fills in operator-specific parameters (like input/output shapes and aggregation dimensions), translating each operator into a standalone SQL query. These queries are then connected according to the computation graph, forming a directed acyclic graph (DAG) of SQL queries.

\para{Post-optimization} \transql applies a \textbf{ROW2COL} transformation to matrix-vector and matrix-matrix multiplication queries. This reduces the cardinality of intermediate results, thereby enhancing \texttt{JOIN} and \texttt{GROUP BY} efficiency. Additionally, consecutive queries are merged into common table expressions (CTEs) to minimize intermediate data materialization and I/O overhead, with natural segmentation occurring at embedding-generation boundaries. Then the final output is generated as a SQL script stored in the local file system.

\para{Execution} When a prompt arrives, an external tokenizer extracts the input tokens and produces an SQL script that inserts those tokens and their corresponding position IDs into the database. Following that, the database then executes sql script generated by \transql recursively, generating tokens one at a time. Each generated token is stored in an output table, which maintains the full decoding history and can be reused in subsequent steps.

\subsection{Model Preprocessing}
\label{sec:preprocessing}

\subsubsection{Chunk-Based Representation}
\label{sec:chunk_representation}
\begin{figure}[t]
    \centering
    \includegraphics[width=\linewidth]{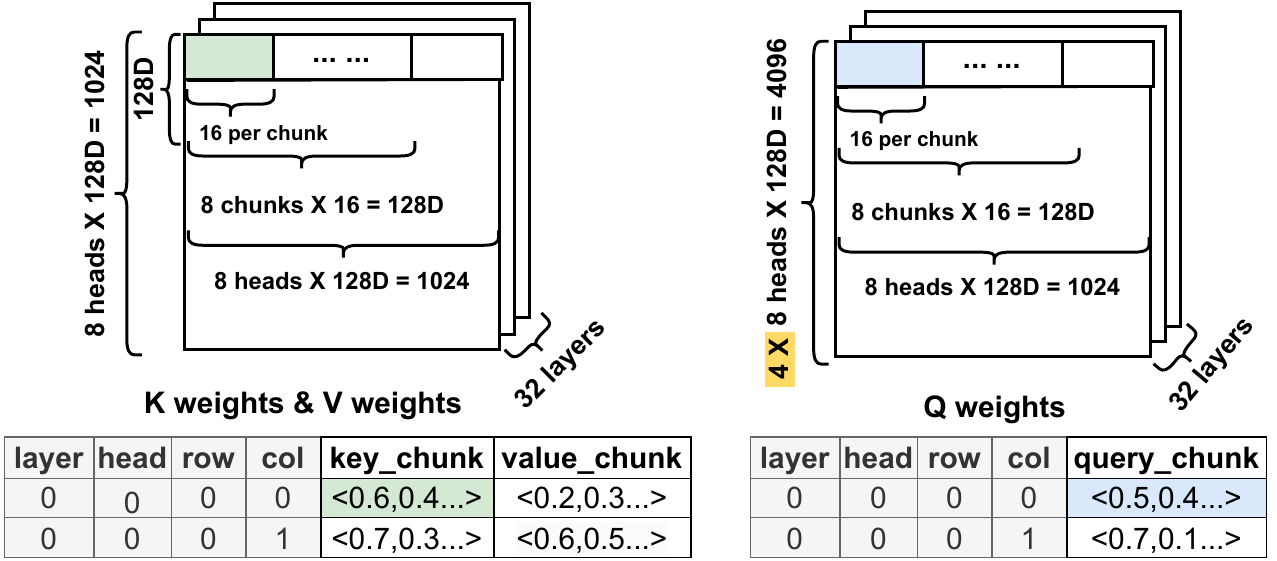}
\caption{Illustration of slicing K, V, Q weights into chunks.}    
\label{fig:chunk_based}
\vspace{-4mm}
\end{figure}

To execute neural operators via SQL queries, we first convert model weight matrices into relational tables. We represent matrices in a \emph{chunked} format, as explored in previous work for in-database LA \cite{declaritive,chunk_matrix1}. Matrices are split into smaller blocks, indexed by row and column. However, because databases rarely offer native matrix support but often support vector operations (e.g., dot products and element-wise arithmetic \cite{vectorized2,clickhouse}), we partition matrices into vectors for broader compatibility. Fig. \ref{fig:chunk_based} illustrates an example of this chunk-based representation.

Specifically, for a large matrix $\mathbf{W} \in \mathbb{R}^{m \times n}$, we split each row into $\lfloor\frac{n}{chunk\_size}\rfloor$ chunks:
\[
    \mathbf{w}_i = 
    \bigl[\mathbf{w}_i^{(0)},\,\mathbf{w}_i^{(1)},\,\ldots,\,\mathbf{w}_i^{(C)}\bigr],
\]
where each chunk $\mathbf{w}_i^{(c)}$ has a fixed dimension determined by a \emph{chunk size} hyperparameter. In the database table, each row thus corresponds to:
\[
    (i,\; c,\; \mathbf{w}_i^{(c)}),
\]
for $0 \leq i < m$ and $0 \leq c < \lfloor\frac{n}{chunk\_size}\rfloor$. This chunking strategy provides flexibility by allowing trade-offs between the number of rows and the workload per thread when performing vector operations.

Algorithm~\ref{alg:chunk} outlines the procedure for converting weight matrices into chunked representations and producing a tuple list suitable for database import. For traditional databases without array support, \transql falls back to a chunk size of 1, storing all matrices in a \texttt{(row\_id, col\_id, value)} schema. Vector operations are rewritten as scalar operations combined with \texttt{GROUP BY}–aggregation patterns in the generated SQL. 

Notably, when a matrix appears as the \emph{second} operand of a multiplication (e.g., \(A\in\mathbb{R}^{m\times k}\) and \(B\in\mathbb{R}^{k\times n}\)), \transql physically transposes \(B\) during preprocessing—storing its column vectors as row vectors—so dot products can be computed directly at inference time without an extra transpose. This alters only the physical layout, not the original computation graph. Our chunking scheme therefore assumes no \emph{algebraic} transposition is required at runtime—an assumption that holds for the transformer models studied here (Llama3 and DeepSeek MoE). We discuss cases that require transposition or re-chunking in Sec.~\ref{sec:specials}.


\begin{algorithm}[t]
\caption{Convert Matrix To Relational Tables}
\label{alg:chunk}
\begin{algorithmic}[1]
\Procedure{ConvertMatrixToChunks}{$matrix$, $chunk\_size$}
    \State $M \gets$ number of rows in $matrix$
    \State $chunkTable \gets \emptyset$
    \For{$i = 0$ to $M-1$}
        \State $row \gets matrix[i]$
        \State $N \gets$ length of $row$
        \For{$c = 0$ to $N-1$ \textbf{step} $chunk\_size$}
            \State $end \gets \min(c + chunk\_size, N)$
            \State $chunk \gets$ slice of $row$ from index $c$ to $end-1$
            \State add tuple $(\text{row\_index}: i, \text{chunk\_index}: c, \text{vector}: chunk)$ to $chunkTable$
        \EndFor
    \EndFor
    \State \Return $chunkTable$
\EndProcedure
\end{algorithmic}
\end{algorithm}


\para{SQL Implementation of Matrix Multiplication}
Within this chunked table structure, we emulate the summation $\sum_{c} a_{ic} b_{cj}$ by joining the chunked rows of the two matrices on their common index $k$. The relational engine then multiplies corresponding entries and aggregates the products to form each entry of the resulting matrix $\mathbf{W}$. Formally, the SQL implementation follows a pattern like:

\begin{small}
\begin{lstlisting}[basicstyle=\ttfamily,mathescape]
SELECT 
    A.$i$ AS $i$, B.$j$ AS $j$, 
    SUM(A.$\mathbf{w}_i^{(c)}$ $\cdot$ B.$\mathbf{w}_j^{(c)}$) AS $w_{ij}$
FROM A JOIN B ON A.$c$ = B.$c$ GROUP BY A.$i$, B.$j$;
\end{lstlisting}
\end{small}

Here, A.$\mathbf{w}_i^{(c)}$ and B.$\mathbf{w}_j^{(c)}$ would each represent chunked slices of the matrices, typically stored as arrays or separate columns. The result is then reorganized (if needed) to match the output dimension $m \times n$. 

This SQL-based matrix multiplication serves as a foundation for implementing other operations (e.g., element-wise activations, softmax) directly in the relational engine, enabling end-to-end inference within a database system.

\subsubsection{Constant Folding} Constants are commonly present in the computation graphs of LLMs, typically serving as parameters for other neural operators. They can take the form of scalars (e.g., token length) or matrices (e.g., weights).  \transql performs \emph{constant folding}: constant tensors and any downstream linear or scalar operations on them are precomputed to avoid redundant work at inference time. In general, for a constant \(A\) in an expression of the form \(\mathrm{Op}(A B) = \mathrm{Op}(A)\, B\), we evaluate \(\mathrm{Op}(A)\) offline. For instance, the attention scaling factor \(\frac{1}{\sqrt{d_k}}\) in Eq. \ref{eq:attn} can be absorbed into the weight \(W_Q\), eliminating the scalar division during inference. This optimization avoids unnecessary computation during SQL execution and simplifies the resulting query plan.

\begin{table*}[t!]
\centering
\caption{Templates for primary operator categories and examples.}
\label{tab:templates}
\resizebox{2\columnwidth}{!}{%
\begin{tabular}{|l|c|l|l|}
\hline
\textbf{Operator Category} & \multicolumn{1}{l|}{\textbf{Operator}} & \textbf{Template} & \textbf{SQL Query} \\ \hline
\multirow{2}{*}{\begin{tabular}[c]{@{}l@{}}Matrix \\ multiplication\end{tabular}} & \multirow{2}{*}{$AB$} & $R_3\leftarrow\Pi_{\mathcal{G},\mathcal{S},\mathrm{DOT}(\mathbf{c_1},\mathbf{c_2})}\bigl(R_1 \;\bowtie_{\mathcal{S}}\; R_2\bigr)$ & \multirow{2}{*}{\begin{tabular}[c]{@{}l@{}}SELECT A.m, B.q, SUM(DOT(A.chunk, B.chunk)) FROM A \\ JOIN B ON A.n = B.p GROUP BY  A.m, B.q\end{tabular}} \\ \cline{3-3}
 &  & $\Pi_{\mathcal{G},\mathrm{SUM(\mathbf{c_3})}}\Bigl(\gamma_{\mathcal{G}}(R_3)\Bigr)$ &  \\ \hline
\begin{tabular}[c]{@{}l@{}}Element-wise \\ function\end{tabular} & \multicolumn{1}{l|}{Sigmoid($A$)} & $\Pi_{\,\mathcal{F},\;f(\mathbf{c_1})}\!\Bigl(R_1\Bigr)$ & SELECT A.m, A.n, 1/(1+exp(-A.chunk)) FROM A \\ \hline
\begin{tabular}[c]{@{}l@{}}Element-wise \\ arithmetic\end{tabular} & $A+B$ & $\Pi_{\,\mathcal{S},\; f(\mathbf{c_1},\mathbf{c_2})}\!\Bigl(R_1 \;\bowtie_{\mathcal{S}}\; R_2\Bigr)$ & \begin{tabular}[c]{@{}l@{}}SELECT A.m, A.n, A.chunk + B.chunk FROM A\\ JOIN B ON A.m = B.p AND A.n = B.q\end{tabular} \\ \hline
Reshape & A.flatten() & $\Pi_{\,\mathcal{G},f(\mathcal{S)},\;\mathbf{c_1}}\!\Bigl(R_1\Bigr)$ & SELECT A.m*M+A.n, A.chunk FROM A \\ \hline
\multirow{4}{*}{Normalization} & \multirow{4}{*}{Softmax(A)} & $R2 \leftarrow \Pi_{\mathcal{F},\mathcal{G},\; \texttt{agg}(f(\mathbf{c_1}))}\!\Bigl(R_1 \Bigr)$ & \multirow{4}{*}{\begin{tabular}[c]{@{}l@{}}WITH exp\_sum AS (SELECT A.m, SUM(SUM(exp(A.chunk))) AS summation \\ FROM A GROUP BY A.m)\\ SELECT A.m, A.n, exp(A.chunk)/summation FROM A\\ JOIN exp\_sum ON A.m = exp\_sum.m\end{tabular}} \\ \cline{3-3}
 &  & $R3 \leftarrow \Pi_{\mathcal{G},\texttt{agg}(\mathbf{c_2})}\gamma_{\mathcal{G}}\!\Bigl(R_2\Bigr)$ &  \\ \cline{3-3}
 &  & \multirow{2}{*}{$\Pi_{\mathcal{F},\mathcal{G},g(\mathbf{c_2},\mathbf{c_3})}\!\Bigl(R_2 \bowtie_{\mathcal{S}} R_3\Bigr)$} &  \\
 &  &  &  \\ \hline
\end{tabular}%
}
\vspace{-3mm}
\end{table*}
\subsection{Template-based SQL Code Generation}
\label{sec:MM_example}

To translate LLM inference computations into executable SQL, we adopt a \emph{template-based SQL code generation} approach.  Each neural operator maps to a predefined SQL template whose placeholders are instantiated with model‑specific parameters—e.g., attention‑head group size (GQA) and token/embedding dimensionality—enabling \transql to flexibly support diverse LLM architectures. This method builds on prior work in structured SQL generation~\cite{sqlizer,ml2sql}, which shows that templates are expressive enough to cover common LA operators.

A major strength of this approach is its \emph{extensibility}: new operators can be supported by simply adding templates. As demonstrated in our evaluation, this flexibility allows \transql to handle both transformer components and convolution kernels.

Table~\ref{tab:templates} summarizes the neural operator categories alongside representative SQL queries. In this section, we focus on the template design for the most important and complex operator categories: matrix multiplication, element-wise functions, and normalization. Additional details for reshape and element-wise arithmetic categories are provided in Appendix.

\subsubsection{Parameterize neural operators}
To match neural operators to their corresponding SQL templates, \transql first identifies the minimal set of parameters required to instantiate a valid SQL query.  For each operator, we record its operand relations, operator type, and index structure (shared dimension vs. free dimensions), then instantiate a parameterized template that compiles to SQL. (i) \emph{Shared dimensions} are indices along which operands interact (multiply, sum, compare). In SQL, they become join keys and aggregation attributes. (ii) \emph{Free dimensions} do not participate in these interactions; in SQL, they appear as grouping or projection columns carried to the output. For example, in $C_{ij}=\sum_{k} A_{ik}B_{kj}$, $k$ is shared (it mediates interaction and is eliminated), while $i$ and $j$ are free, becoming the row and column indices of $C$.

During preprocessing, \transql categorizes all neural operators in the inference computation graph into the following types: (i) element-wise functions (e.g., \texttt{Sigmoid}, \texttt{SiLU}), (ii) element-wise arithmetic (e.g., vector addition, element-wise multiplication), (iii) matrix–matrix and matrix–vector multiplication, (iv) shape manipulation (e.g., \texttt{view}, \texttt{reshape}), and (v) normalization-related operations (e.g., \texttt{RMSNorm}, \texttt{Softmax}). After categorization, raw tensor operands' shared dimensions are transformed into chunk-based representations and subsequently converted into relational tables, as detailed in Sec.~\ref{sec:preprocessing}.

Formally, a neural operator acting on $p$ operands is represented as:
\[
\operatorname{OP}_{\texttt{attr}}\Bigl(\{R_1, \dots, R_p\},\, \mathcal{F},\, \mathcal{S},\, \mathcal{G}\Bigr),
\]
where $\mathcal{F}$ denotes the set of \emph{free dimensions} retained in the output for each operand $R_i$, and $\mathcal{G}$ represents the set of dimensions to be used in \texttt{GROUP BY} clauses in the SQL template. $\mathcal{S}$ encodes the mappings of \emph{shared dimensions} between operands, which are further translated into SQL join keys. The subscript $\texttt{attr}$ indicates a set of operator-specific attributes that guide the final projection logic, such as function types, reshaping rules, or normalization strategies.


\smallskip

\para{Matrix-Matrix Multiplication} Matrix multiplication \(\mathbf{X} = \mathbf{A}\mathbf{B}\), where \(\mathbf{A}\in\mathbb{R}^{m\times r}\) and \(\mathbf{B}\in\mathbb{R}^{r\times n}\), can be calculated through:
\[
    x_{ij}=\sum_{k=0}^{\lfloor\frac{r}{c}\rfloor} \mathbf{a}_{ik}\cdot\mathbf{b}_{kj}, \quad \text{for } i\in[0,m), j\in[0,n),
\]
where free dimensions $m$ and $n$ and shared dimension \(r\) (subdivided into chunks by $c$) are identified. Matrix multiplication is parameterized as:
\begin{equation}
    \operatorname{Matmul}\Bigl(\{A, B\},\, \varnothing,\, \{(r_A, r_B)\},\, \{m_A, n_B\}\Bigr),
\end{equation}
where $A$ and $B$ are the input matrices, $\{m_A, n_B\}$ are the \emph{group-by dimensions} corresponding to the output row and column indices, and $\{(r_A, r_B)\}$ represents the \emph{shared dimensions} over which the summation is performed. When input data is imported into the database, the right-hand operand in matrix multiplication is transposed to convert its column vectors into row vectors, aligning with how dot products are computed in databases. Note that the symbolic notation used here abstracts away from the actual physical relational schema.

Since the summation in matrix multiplication translates to a \texttt{GROUP BY + SUM} operation in SQL, both $m_A$ and $n_B$ are included in the set $\mathcal{G}$ to ensure the correct grouping of partial products in the SQL query.

\para{Element-wise functions} Element-wise functions apply a transformation independently to each element of matrices using a function $f$. A representative class of such operations includes activation functions such as \texttt{Sigmoid}, \texttt{ReLU}, and \texttt{SiLU}. Since these operations do not involve any interactions across elements—i.e., no joins or aggregations are required—all dimensions are considered \emph{free dimensions}.
For example, $\operatorname{Sigmoid}(A)$, where $A\in\mathbb{R}^{m\times n}$, can be parameterized as:
\[
\operatorname{Sigmoid}\Bigl(\{A\},\,\{m_A,n_A\},\,\varnothing,\,\varnothing\Bigr).
\]
Because Sigmoid function applies to elements in matrix $A$ independently, the shared dimension here is an empty set.

\para{Normalizations}
Normalization-like operations are ubiquitous in LLM architectures. These operators typically follow a two-step pattern: (i) apply a function to each element, (ii) apply aggregation in chunks as well as over a shared dimension, and (iii) apply a function to each element and aggregated values over the shared dimension. This general pattern not only includes statistical normalizations (e.g., \texttt{layernorm}), but also functions like \texttt{softmax}. 

To capture the structure of normalization operations, we parameterize them using three components: an element-wise function $f$, an aggregation function $\mathit{agg}$, and a final element-wise post-processing function $g$. These functions define the sequence of transformations in the SQL template: applying $f$ to the input, aggregating with $\mathit{agg}$ over the shared, and then applying $g$ to normalize each element using the aggregated result. We illustrate this with two representative examples: \texttt{softmax} and \texttt{layernorm}. 

\texttt{Softmax} is typically applied row-wise, transforming a vector $\mathbf{a}_i$ from a matrix $A \in \mathbb{R}^{m \times n}$ as:
\[
\operatorname{Softmax}(\mathbf{a}_i) = \frac{\exp(a_{ij})}{\sum_{j=0}^{n-1} \exp(a_{ij})}.
\]
This operation first applies $\exp$ element-wise, computes a sum over the shared dimension (i.e., across each row), and finally normalizes each element by dividing by the corresponding aggregated sum.

We express this using the following parameterization:
\[
\operatorname{Normalize}_{\exp,\,\texttt{SUM},\,\texttt{div}}\left(A,\, \{n_A\},\, \{(m_A,m_A)\},\,\{m_A\}\right),
\]
where $\exp$ is the first element-wise function, \texttt{SUM} is the aggregation over $\{n_A\}$, and \texttt{div} performs the final normalization step.

Similarly, \texttt{layernorm} can be represented in the same framework by using the \texttt{identity} function as the first step (i.e., $f = \texttt{id}$), which leaves the input unchanged before aggregation. 

Furthermore, building on the \texttt{layernorm} pattern, we can perform global aggregation across each row of the input matrix by modifying the parameterization: specifically, by setting the final function $g$ to the \texttt{identity} function and assigning all dimensions as shared dimensions. For example, this allows us to compute the maximum value in each row through a single aggregation step. These variants highlight the expressiveness and flexibility of our parameterization framework in capturing a wide range of normalization and reduction behaviors within a unified abstraction.

\subsubsection{Special Cases}
\label{sec:specials}
Some neural operators fall outside the five core categories discussed earlier. While these cases are relatively infrequent, \transql handles them individually due to their unique computational semantics.

\para{Lookup Table} Lookup tables are commonly used in the input layer to retrieve token embeddings and in the output layer to map final activations to vocabulary tokens. These operations can be efficiently implemented using a simple equi-join on token strings or token IDs. For instance, to retrieve embeddings for a sequence of tokens stored in an \texttt{input\_token\_table}, we can join it with an \texttt{embedding\_table} that maps tokens to their corresponding vector representations:

\begin{lstlisting}[basicstyle=\ttfamily\small]
SELECT 
    input.token_id, embed.chunk_id, embed.chunk
FROM input_token_table AS input
JOIN embedding_table AS embed
ON input.token = embed.token
\end{lstlisting}

Here, \texttt{row\_id} maintains the original token order, and the join retrieves the matching embedding vector for each token. This pattern supports both input embedding lookup and final output decoding, and integrates naturally into the relational execution framework.

\para{Transpose} The transpose operation is non-trivial under a chunk-based representation, as it requires unnesting and reassembling all values with swapped dimension indices. 
In the models discussed here, no runtime transposition is needed. Weight matrices are static, and intermediate matrices are consumed only by specific, fixed operations. Thus, if a transposition appears in the original computation graph, we can instead rewrite the operation by pre‑transposing the corresponding weight matrix. For example, consider a sequence of matrix multiplications:
\[
X = A \cdot (B C^\top)^\top,
\]
where $A$, $B$, and $C$ are input matrices. A naive evaluation would compute $B C^\top$, then transpose the intermediate result, and finally multiply with $A$. However, we can algebraically rewrite this as:
\[
X^\top =  B C^\top A^\top,
\]
which avoids the intermediate transpose. To enable this transformation, we store $A^\top$ in the database during model import instead of $A$, effectively bypassing the need for transposition during inference.

If intermediate transposes are required for other model types, \transql can incorporate them purely in SQL. Starting from a chunked layout \texttt{(row\_id, col\_chunk, vec)}, we (i) unnest each vector into tuples \texttt{(row\_id, col\_id, value)} with \texttt{col\_id = col\_chunk * chunk\_size + offset}; (ii) swap indices to \texttt{(col\_id, row\_id, value)}; and (iii) re-aggregate by \texttt{col\_id} into chunked vectors, yielding the transposed table. This incurs some overhead but does not compromise the completeness or correctness of our approach.

\subsubsection{SQL code generation}
\label{sec:code_generation}

Using the parameterization results, we populate predefined SQL templates to generate queries, each forming a temporary view with a unique identifier. Most of these views are later merged during post-optimization via \texttt{WITH}-clause rewriting, as detailed in the next section.

In Table~\ref{tab:templates}, we follow standard relational algebra notations: $\Pi_{*}$ denotes a \emph{projection}, where $*$ indicates either a set of attributes or functions applied to them; $\bowtie_{*}$ represents a \emph{join}, with $*$ specifying an equi-join condition (or a cross join if omitted); and $\gamma$ denotes a \emph{group-by} operation, typically followed by a projection applying an aggregation function.

We assume input tensors with dimensions $A \in \mathbb{R}^{m \times n}$ and $B \in \mathbb{R}^{p \times q}$. In their chunk-based relational representation, $c_A$ and $c_B$ refer to the chunk identifiers for $A$ and $B$, respectively. The functions applied to chunks are vector operations supported by databases.

The five core categories and special cases in our framework cover the neural operators used in transformer‑based LLMs (e.g., LLaMA, QWen, DeepSeek MoE).  With the templates presented here, we already support both dense and MoE models in this family. While not exhaustive, SQL’s expressiveness and our modular template design allow easy integration of new operators through additional templates and parameter mappings. By extending our templates and implementing parameterization accordingly, \transql can support broader model families. For convolutional networks, 2D convolutions can be rewritten via the common \emph{im2col} trick into matrix multiplications and thus reuse our MatMul template; pooling layers can be expressed by generating window coordinates (via row/column indices) followed by a \texttt{GROUP BY}–\texttt{MAX} (or \texttt{AVG}) aggregation, requiring only minor template additions. The main change is the input format (image tensors instead of token sequences); once ingested, the workflow in Fig.~\ref{fig:workflow} remains unchanged. Crucially, these adaptations are one-time template extensions—no modifications to the execution environment—demonstrating \transql’s extensibility and portability.

\section{Post-optimization}
\label{sec:opt}
Although the SQL templates translate neural operators into view creating queries, end‑to‑end performance can be suboptimal: repeated materialization incurs heavy I/O, and large intermediates, especially from matrix multiplication, slow \texttt{JOIN}/\texttt{GROUP BY} and constrain parallelism. We address this with three optimizations that reduce I/O workload and expose more parallel work on vectorized engines: \emph{temporary view elimination}, \emph{table fusion}, and \emph{ROW2COL} pivoting. Together, these changes better exploit the planner and execution engine and substantially improve performance.

\subsection{Temporary View Elimination}
After generating SQL templates for individual neural operators, the next step is to assemble these templates following the original computation graph. This process initially produces many SQL subqueries and temporary views, which can cause significant I/O overhead. To address this, we leverage Common Table Expressions (CTEs) to merge multiple subqueries into a single, unified SQL statement. The optimization procedure is summarized in Algorithm~\ref{alg:temp_view_elimination}.

The computation graph is first topologically sorted (line 4) to ensure correct execution order. Each node is then visited to determine whether it can be merged into the preceding CTE block. If a node is identified as a \emph{critical node}, the currently accumulated nodes are materialized into a table (lines 5–12); otherwise, the node is absorbed into the ongoing CTE expression. A critical node is one whose output (i) is consumed by multiple downstream operators or (ii) is subsequently modified by one or more operators. To avoid recomputation, we materialize such outputs as tables. In transformer-based LLMs, critical nodes fall into two categories: 
\smallskip

\noindent\textbf{Key and Value Vectors of Attention Heads:} These vectors are essential for constructing the key–value cache \cite{kvcache} used in subsequent decoding steps. Materializing them allows efficient reuse during generation without recomputing the attention mechanism.

\noindent\textbf{Embeddings for Residual Connections:} In models with residual connections, the original input embeddings must be preserved for later addition to the updated embeddings, as illustrated in Fig.~\ref{fig:moe_vs_dense} (marked as \textit{skip}). Materializing these embeddings ensures that the residual paths are correctly maintained during inference.

\smallskip

In the final step, any remaining unvisited nodes are folded into the last CTE, completing the SQL script; this optimization cuts the number of materialized intermediates per layer for Llama3 8B from 38 to 7, substantially reducing I/O from repeated materialization and scans.

\begin{algorithm}[t]
\caption{Temporary View Elimination}
\label{alg:temp_view_elimination}
\begin{algorithmic}[1]
\Procedure{OptimizeCTEQueries}{nodes, graph}
    \State $finalSQL \gets []$ \Comment{List of final SQL statements}
    \State $cteChain \gets []$ \Comment{Cache of SQL fragments for merging via CTEs}
    \State $topoNodes \gets$ \textsc{TopologicalSort}$(graph)$ \Comment{Obtain nodes in topological order}
    \ForAll {$node$ in $topoNodes$}
        \If{\textsc{IsCriticalNode}($node$)} \Comment{Determine if node requires materialization}
            \If {$cteChain \neq []$}
                \State $cteQuery \gets$ \textsc{MergeCTEs}$(cteChain)$ \Comment{Merge non-critical nodes into one CTE query}
                \State Append $cteQuery$ to $finalSQL$
                \State Clear $cteChain$
            \EndIf
            \State Append $node.SQL\_query$ to $finalSQL$ \Comment{Materialize critical node separately}
        \Else
            \State Append $node.SQL\_query$ to $cteChain$ \Comment{Accumulate non-critical node query}
        \EndIf
    \EndFor
    \If {$cteChain \neq []$}
        \State $cteQuery \gets$ \textsc{MergeCTEs}$(cteChain)$
        \State Append $cteQuery$ to $finalSQL$
    \EndIf
    \State \Return \textsc{CombineSQLStatements}($finalSQL$) \Comment{Combine all SQL statements into a final script}
\EndProcedure
\end{algorithmic}
\end{algorithm}

\subsection{Table Fusion}

In LLM inference, the \emph{query}, \emph{key}, and \emph{value} vectors are essential intermediate results used to compute attention scores for each token. These vectors are obtained by projecting the input embeddings through their respective weight matrices:
\[
Q = \text{embedding} \times W_Q,\,
K = \text{embedding} \times W_K,\,
V = \text{embedding} \times W_V,
\]
where \( W_Q, W_K, W_V \in \mathbb{R}^{d_{\text{model}} \times d_k} \) are the projection weights.

Although computed independently in the model’s computation graph, these projections share the same input and produce outputs with identical schema and cardinality. Naively materializing them separately results in redundant scans of the embedding table and generates three temporary output tables, breaking the continuity of the CTE pipeline and increasing I/O overhead. Moreover, since these vectors are consumed together in a single attention layer, querying them independently prevents the query optimizer from performing fine-grained parallelization, as it can only schedule at the table level.

To address this inefficiency, we adopt a fusion-based optimization by concatenating the projection weights into a single extended matrix:
\[
[Q\;\|\;K\;\|\;V] = \text{embedding} \times W_{QKV},\quad \text{where } W_{QKV} \in \mathbb{R}^{d_{\text{model}} \times 3d_k}.
\]

\begin{figure}[t]
    \centering
    \includegraphics[width=0.85\linewidth]{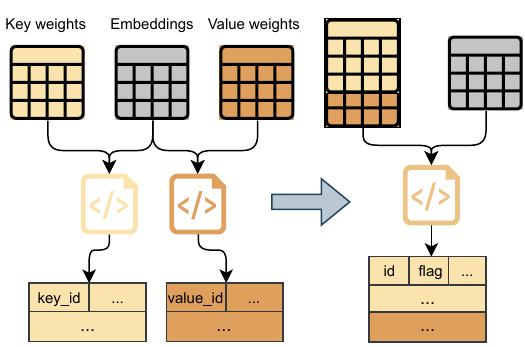}
\caption{Merge computation of key and value vectors into a single CTE.}    
\label{fig:cte_fusion}
\vspace{-5mm}
\end{figure}

In \transql, this is implemented as a table fusion optimization. The weight tables for \( W_Q \), \( W_K \), and \( W_V \) are vertically merged using \texttt{UNION ALL}, with a flag column distinguishing the projection type. During SQL generation, this flag becomes part of the shared dimensions, allowing all three vectors to be computed in a single SQL query. Figure~\ref{fig:cte_fusion} illustrates this optimization.

 Merging Q/K/V weight tables into one single relation lets us scan the weights once and compute all three matrix multiplications in a single MatMul SQL query, distinguished only by different \texttt{GROUP BY} keys. Since vectorized engines load contiguous column chunks into SIMD registers, a unified table lets the executor stream Q/K/V columns together, avoiding cache flushes and repeated scans while exposing finer-grained parallelism (each column block can be pipelined across cores). The same idea applies to SwiGLU feed-forward layers: the two parallel projections can be fused into one table, enabling a single pass that computes both branches and eliminating redundant joins and I/O.

\subsection{ROW2COL Pivoting}

Although SQL is expressive for neural operators, executing them directly in relational engines can be inefficient because join‑based alignment inflates intermediate results, forces key materialization, and limits intra‑operator parallelism. In matrix multiplication $X=AB$, tensor frameworks (NumPy~\cite{numpy}, PyTorch~\cite{pytorch}) access rows and columns by array index, whereas relational systems align vectors via scans and equi‑joins—as illustrated by the matrix‑multiplication example in Sec.~\ref{sec:chunk_representation}—incurring additional overhead.

 Unlike tensor frameworks that operate on in-memory dense matrices and traditional \emph{row-oriented} RDBMSs, modern analytical engines (e.g., DuckDB, ClickHouse, Greenplum) use columnar storage to exploit parallelism across independent columns and to improve cache locality on contiguous column values. Building on the chunked formulation of matrix multiplication—where chunk \(k\) of \(A\) pairs one-to-one with chunk \(k\) of \(B\)—we introduce \textsc{ROW2COL}, a logical schema rewrite that pivots chunk rows into columns. ROW2COL reduces join cardinality and scanned rows and exposes contiguous, independent column slices for parallel processing. The transformation uses \texttt{PIVOT} where available (e.g., SQL Server, DuckDB) or conditional aggregation with \texttt{CASE} otherwise.


For simplicity, we use DuckDB built-in \texttt{PIVOT} syntax as an example. Assume $A$ and $B$ are chunked into 64 vectors. The original schema is \(\texttt{(row\_id, chunk\_id, vec)}\). ROW2COL pivots this to
\(\texttt{(row\_id, chunk0, \ldots, chunk63)}\), so each chunk becomes a dedicated column. Based on the new schema, ROW2COL rewrites the matrix multiplication query in Sec. \ref{sec:chunk_representation} as:
\begin{lstlisting}[basicstyle=\ttfamily\small]
WITH c0 AS (SELECT A.row_id, B.row_id, 
        DOT(A.chunk0, B.chunk0) AS v0
    FROM A_pivot CROSS JOIN B_pivot 
    ORDER BY A.row_id, B.row_id),
    ...
c63 AS (SELECT A.row_id, B.row_id, 
        DOT(A.chunk63, B.chunk63) AS v63
    FROM A_pivot CROSS JOIN B_pivot 
    ORDER BY A.row_id, B.row_id)
SELECT A.row_id AS i, B.row_id AS j, 
    v0+...+v63 AS value_ij
FROM c0 POSITIONAL JOIN c1
\end{lstlisting}
 This optimized query yields multiple benefits: each CTE subquery computes the dot product for a single chunk pair, enabling independent, parallel execution; the final step replaces a costly \texttt{Equi-Join} with a \texttt{POSITIONAL JOIN} that aligns rows by order, eliminating explicit key matching. Overall, \emph{ROW2COL} (i) cuts the number of source rows and the cardinality of intermediate joins, accelerating scans and join processing; (ii) exploits modern vectorized engines \cite{vectorized1,vectorized2,vectorwise} by treating columns as contiguous arrays, reducing function-call overhead and improving cache locality; and (iii) exposes column-aligned chunks that can be processed in parallel, increasing opportunities for the optimizer to schedule concurrent work.

Nevertheless, \textsc{ROW2COL} is not a silver bullet. Although DuckDB~\cite{vectorized2} and ClickHouse~\cite{clickhouse} support \texttt{POSITIONAL JOIN}, it is non‑standard and often unavailable elsewhere, so fallbacks such as \texttt{SORT–MERGE JOIN} are needed; these can perform comparably when subqueries are consistently pre‑ordered. Pivoting also changes the schema and can introduce extra  materialization, and in some cases this overhead outweighs the reduced join costs. Very wide pivots can strain planning and execution, especially once the column count exceeds effective parallelism. We empirically characterize these trade‑offs in Section~\ref{sec:perf_insights} under varying hardware settings.

 We therefore apply \textsc{ROW2COL} conservatively: only when the operator is chunk-parallel with a moderate number of chunks (e.g., embeddings or K/V tables), the widened schema remains within the engine’s practical limits (vector width, planner constraints), and the anticipated drop in join cardinality outweighs the pivoting cost (heuristically estimated from chunk count, and vector width). We skip ROW2COL for extremely wide pivots  or very large chunks (CPU saturation). A full cost model would weigh pivot cost, expected cardinality reduction, width relative to core count, and cache pressure. Designing such a comprehensive cost model requires substantial effort, which we leave for future work.



\section{Evaluation}
\label{sec:experiments}
In this section, we comprehensively evaluate the performance, efficiency, and generality of \transql as a lightweight SQL-based inference engine for large language models and deep learning operators. Our evaluation is designed to answer the following questions: 
\\ \textbf{Q1}: How does \transql compare to state-of-the-art inference frameworks regarding latency under resource-constrained settings? 
\\ \textbf{Q2}: What is the contribution of post-optimization techniques to overall performance? 
\\ \textbf{Q3}: How does memory capacity impact execution efficiency? 
\\ \textbf{Q4}: Can the same SQL-based framework be extended to support other types of neural operators, such as convolutions?

We evaluate both dense and MoE LLMs using real-world models (Llama3 and DeepSeek MoE), run experiments on low-memory hardware, and compare against two widely-used inference frameworks—DeepSpeed and Llama.cpp. Additionally, we benchmark \transql against DL2SQL with convolution kernels to demonstrate extensibility.

\subsection{Experimental Setup}

\subsubsection{Models}
We evaluate \transql on two representative LLMs: Llama3 8B (16.1GB), a dense transformer with GQA and SwiGLU, and DeepSeek MoE \cite{deepseekmoe} (32.8GB), a mixture-of-experts model with 6 expert activations for each token. These models reflect the core architectural patterns of many popular LLMs (e.g., QWen \cite{qwen2}, Mixtral \cite{mixtral}) and cover the key neural operators addressed by our SQL templates. The models used here are unquantized full-precision, as relational databases lack native support for quantized arithmetic. Compression methods are orthogonal to this research and can be applied separately.

Prompt tokens are pre-generated, and each run at a given length uses the same prompt. We restart the database before every run to clear caches. Prompt lengths follow the LMSys-Chat \cite{lmsys-chat} average of 69.3 tokens, so we test 25–200 tokens to cover typical conversations. Although very long prompts are practical in some settings, they are relatively uncommon on resource‑constrained devices acting as personal assistants. Evaluating up to 200 tokens is sufficient to reveal \transql’s advantages and limitations.

\subsubsection{Hardware and Database}
We conduct experiments on an AWS \texttt{c7.2xlarge} instance with  4 physical CPU cores and 16GB RAM—comparable to typical personal devices. Our focus is on evaluating \transql under memory-constrained conditions, not on edge-specific CPU architectures; since all baselines run on the same hardware, comparisons remain fair. To simulate tighter memory limits, we also evaluate performance under reduced RAM in Sec.~\ref{eval:mem}.

We use DuckDB \cite{vectorized2} as our backend, a lightweight analytical database supporting out-of-core execution, vectorized processing, and array operations—key features for expressing neural computations in SQL. To demonstrate \transql’s compatibility with other databases, we also evaluate its performance on ClickHouse \cite{clickhouse}.

\subsubsection{Baselines}
We compare \transql with two production-grade inference frameworks: DeepSpeed Inference and Llama.cpp. 
\\(I) \emph{DeepSpeed} is a transformer inference engine developed by Microsoft that includes support for tensor parallelism, kernel fusion, and NVMe-based weight offloading. It is widely used in production-level deployments.
\\(II) \emph{Llama.cpp} is a lightweight LLM serving framework. It enables out-of-core execution via memory-mapped weights and is especially popular for edge deployments and personal devices. 

Several research efforts have explored model offloading to disk; however, many of these approaches either require an additional training step to summarize model weights—effectively producing a new model \cite{powerinfer,llmflash}—or are specifically tailored for GPU-based execution and memory management, limiting their generalizability \cite{flexgen}. Since our goal is to explore a broadly applicable method for deploying LLMs on personal or edge devices, these approaches are not directly comparable to \transql. Instead, we select \texttt{DeepSpeed} and \texttt{Llama.cpp} as baselines due to their wide adoption and contrasting design goals: \texttt{DeepSpeed} targets both high-performance training and inference, while \texttt{Llama.cpp} focuses on inference on various hardware settings. Crucially, both support inference with model sizes that exceed physical memory, aligning well with our deployment setting.



\subsubsection{Metrics}
We report performance using two key metrics that correspond to the typical stages of LLM inference: prefill latency and decoding latency. Prefill latency measures the time required to process the input prompt and produce the first output token. This stage is the most computationally intensive, as it involves full attention and feed-forward passes over the entire prompt. Decoding latency measures the average time taken to generate each subsequent token, assuming cached key and value vectors. 
\begin{figure}[!t]
    \centering
    \includegraphics[width=\linewidth]{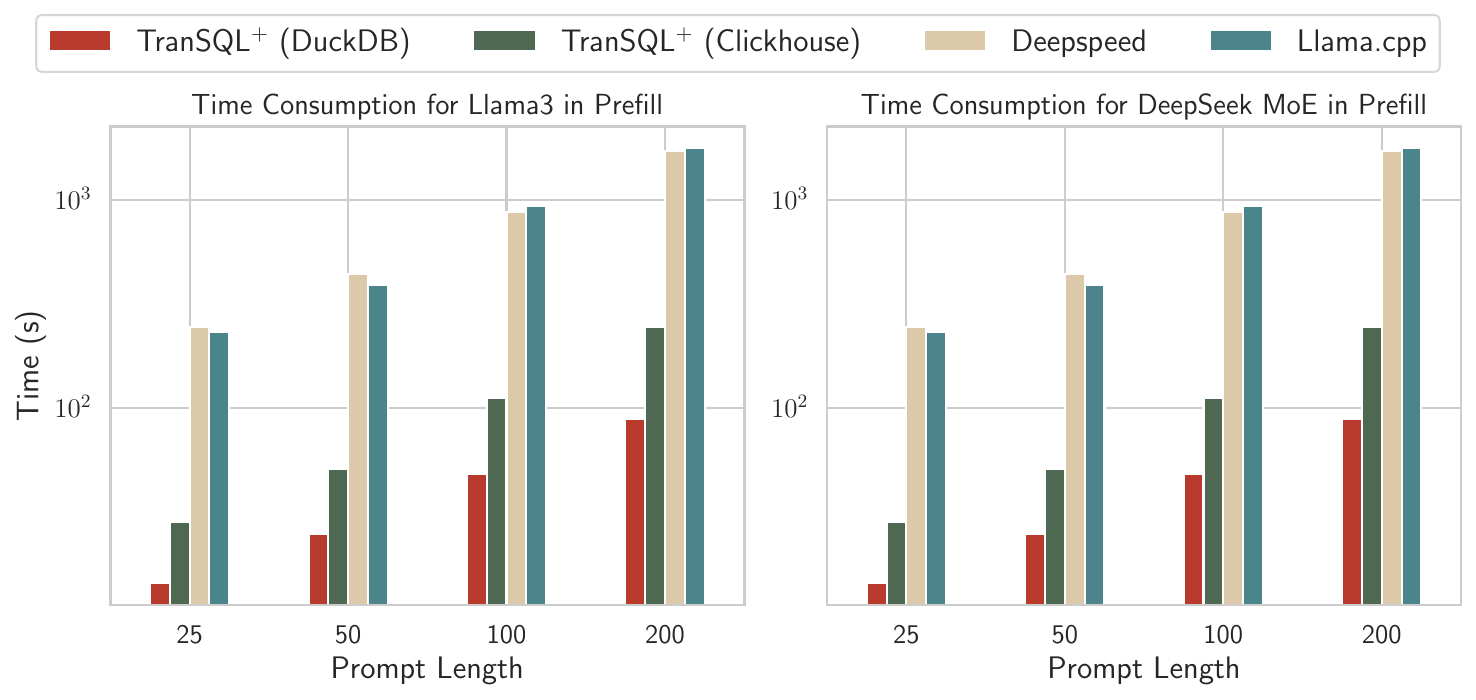}
\caption{\textbf{R1.O3, R2.O15:} Prefill latency with varying prompt length. \transql gains up to 20x speedups compared to Deepspeed and Llama.cpp.}
\label{fig:time_prefill}
\vspace{-2mm}
\end{figure}

\begin{figure}[!t]
    \centering
    \includegraphics[width=\linewidth]{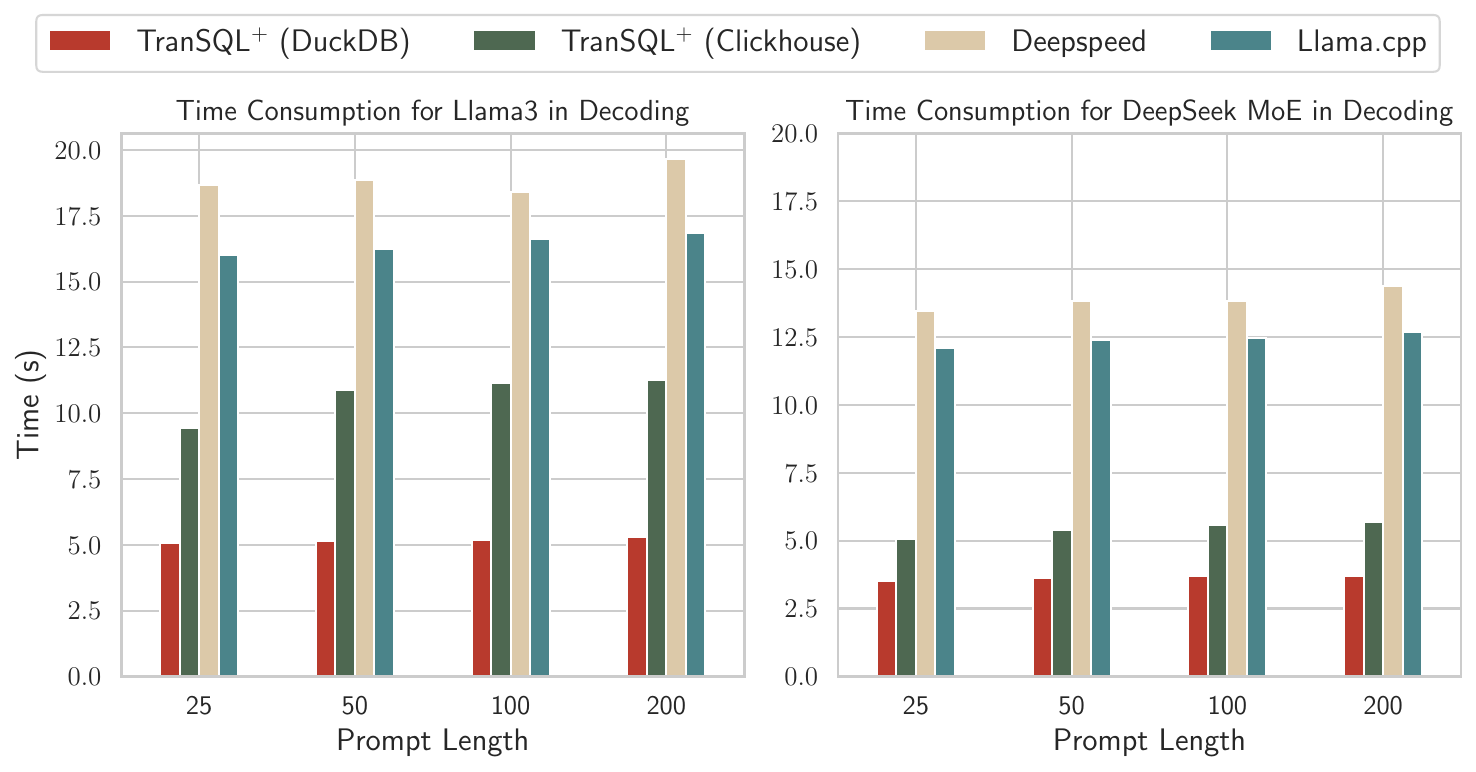}
\caption{Decoding latency with varying prompt length. \transql gains up to 4x speedups compared to Deepspeed and Llama.cpp.}
\label{fig:time_decoding}
\vspace{-5mm}
\end{figure}

\subsection{Overall Latency Evaluation (Q1)}
We evaluate the end-to-end inference performance of \transql \emph{with all post-optimizations} using two large language models—Llama3 8B and DeepSeek MoE—across a range of input prompt lengths. We report the latency for both the prefill phase (i.e., time to generate the first token) and the decoding phase (i.e., average time per generated token after the first). 

\para{Prefill Latency}
Fig.~\ref{fig:time_prefill} summarizes the prefill latency. In the Llama3 and DeepSeek MoE models,  \transql\ on DuckDB and ClickHouse consistently achieves lower prefill latency than both deep learning serving frameworks across all prompt lengths. Moreover, \transql\ on DuckDB is faster than on ClickHouse, indicating that performance depends on the underlying engine. As prompt lengths increase from 25 to 200 tokens, the latency of all baselines grows, but \transql demonstrates significantly better scalability. This is especially evident in the MoE model, where the benefits of our query optimizations—such as \texttt{ROW2COL} pivoting and selective materialization—are amplified due to the model’s partially activated pattern. 

\para{Decoding Latency}
During decoding, where only one token is generated at a time, the performance gap between methods narrows but remains consistent, as shown in Fig.~\ref{fig:time_decoding}. \transql continues to outperform all other approaches on both models.  The gap between DuckDB and Clickhouse also narrows due to less computation and I/O workload. In the dense Llama3 model, decoding times remain relatively stable across prompt lengths. 

\begin{figure}[t]
    \centering
    \includegraphics[width=0.9\linewidth]{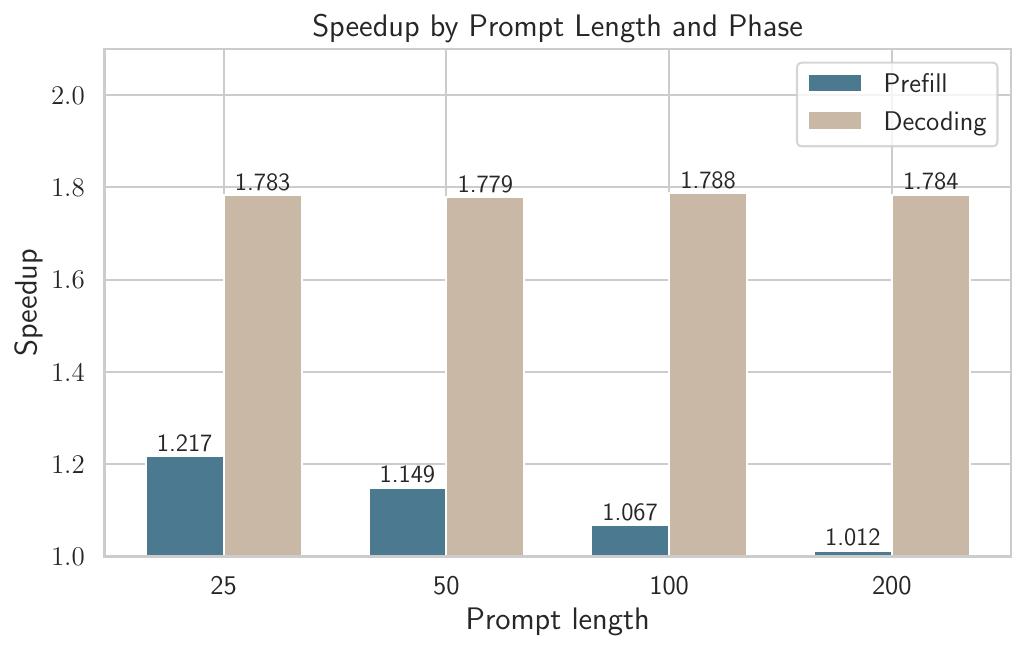}
\caption{ Speedup of Prefill with index vs. without index on Experts for DeepSeek MoE}
\label{fig:speedup_index}
\vspace{-3mm}
\end{figure}

{\para{Effectiveness of Index for DeepSeek MoE}
 Database indices accelerate MoE inference by exploiting low selectivity—only a small subset of experts is active per token. Fig. \ref{fig:speedup_index} shows the speedup from indexing the expert layer in DeepSeek MoE. During decoding, indexing yields a nearly stable 1.78× speedup, since only 6 of 64 experts are queried, reducing the table scan time. In the prefill stage, longer prompts activate more experts, causing the computation to resemble dense inference (as in Llama3). Consequently, by 200 tokens, the index provides negligible benefit.

\para{Performance on GPUs}  For reference, we also ran Llama3 8B on a single A100 GPU (with sufficient memory to hold all weights) using PyTorch. Prefill latency for a 200‑token prompt was approximately 60ms, and per‑token decoding latency was approximately 10ms—far below any baselines in our study. GPUs are, unsurprisingly, the optimal platform when high-end hardware is available. Our goal, however, is not to replace such infrastructure but to enable practical inference under strict resource constraints; TranSQL+ targets precisely these low-memory, heterogeneous environments where a GPU is unavailable or impractical.

\para{Summary}
Overall, these results show  that \transql outperforms existing inference frameworks on resource‑constrained hardware. Its template-based SQL generation plus relational optimization delivers competitive speed without hardware-specific engineering or external runtimes, showing that databases can serve as practical, portable LLM engines. Unless otherwise noted, we use DuckDB as the backend in the remaining experiments.




\subsection{Performance Insights (Q2)}
\label{sec:perf_insights}
To assess the efficiency gains from our post‑optimizations, we (i) provide an operator‑level performance breakdown to show where time is saved, (ii) run ablations to isolate each optimization’s contribution, and (iii) explore ROW2COL chunking parameters to quantify sensitivity and guide future tuning. All experiments rely solely on the database’s native behavior (e.g., vectorized execution by default); we do not enable any special features beyond defaults.

\begin{figure}[t]
    \centering
    \includegraphics[width=\linewidth]{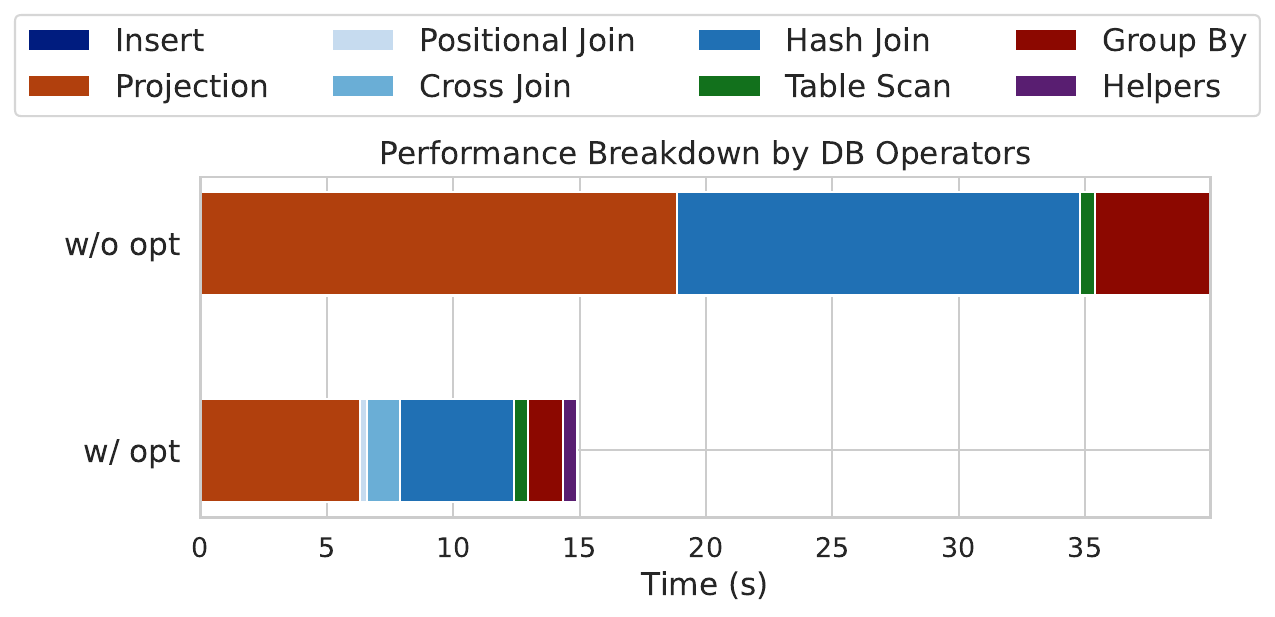}
\caption{Join time drops by over 60\% because the optimizations reduce intermediate cardinality. “w/ opt” denotes \transql with all three optimizations; “w/o opt” denotes \transql without any optimizations.}
\label{fig:performance_breakdown}
\vspace{-3mm}
\end{figure}
\para{Performance Breakdown by Database Operations}
Fig. \ref{fig:performance_breakdown} shows that our post-optimization strategy reduces total execution time by 64.8\% compared to unoptimized SQL queries. Optimizations we discussed in Sec. \ref{sec:opt} aim to fully utilize parallelization and reduce cardinality of intermediate results, particularly in hash joins and group-by aggregations. The operator-level breakdown confirms the impact of these optimizations: join time is reduced by 62.1\%, and group-by operations see a 70.3\% reduction. Moreover, our use of optimized CTEs allows the database’s vectorized execution engine to exploit greater parallelism. This results in a 66.4\% reduction in projection time, highlighting improved cache locality and reduced function call overhead. 



\para{Ablation Study for Post-optimizations}
 We evaluate the individual and combined impact of our three post optimizations (i.e. temp view elimination (denoted as CTE), table fusion, and ROW2COL) in Fig.~\ref{fig:ablation}. Each optimization outperforms a direct SQL translation baseline. In prefill, applying all three together yields the greatest speedup for the Llama3 model, whereas the MoE model see smaller gains: longer prompts activate more experts, incurring I/O overhead that these optimizations cannot fully eliminate. During decoding, MoE benefits more because only a few experts are accessed per token, allowing our optimizations to effectively exploit parallelism of the vectorized execution engine.

Among the three optimizations, CTE provides only modest improvements, since modern databases already lazily evaluate temporary views; materialized CTEs simply avoid repeated schema creation and view re‑evaluation. Table fusion delivers noticeable speedups in attention layers by consolidating Q/K/V weight tables into a single relation, thereby reducing join overhead. ROW2COL has the most dramatic effect in both stages by exploiting columnar storage and SIMD‑parallel operations to minimize scan volume. Together, these targeted optimizations leverage the strengths of vectorized, columnar execution engines to automatically accelerate SQL codes generated by \transql.

\begin{figure}[t]
    \centering
    \includegraphics[width=\linewidth]{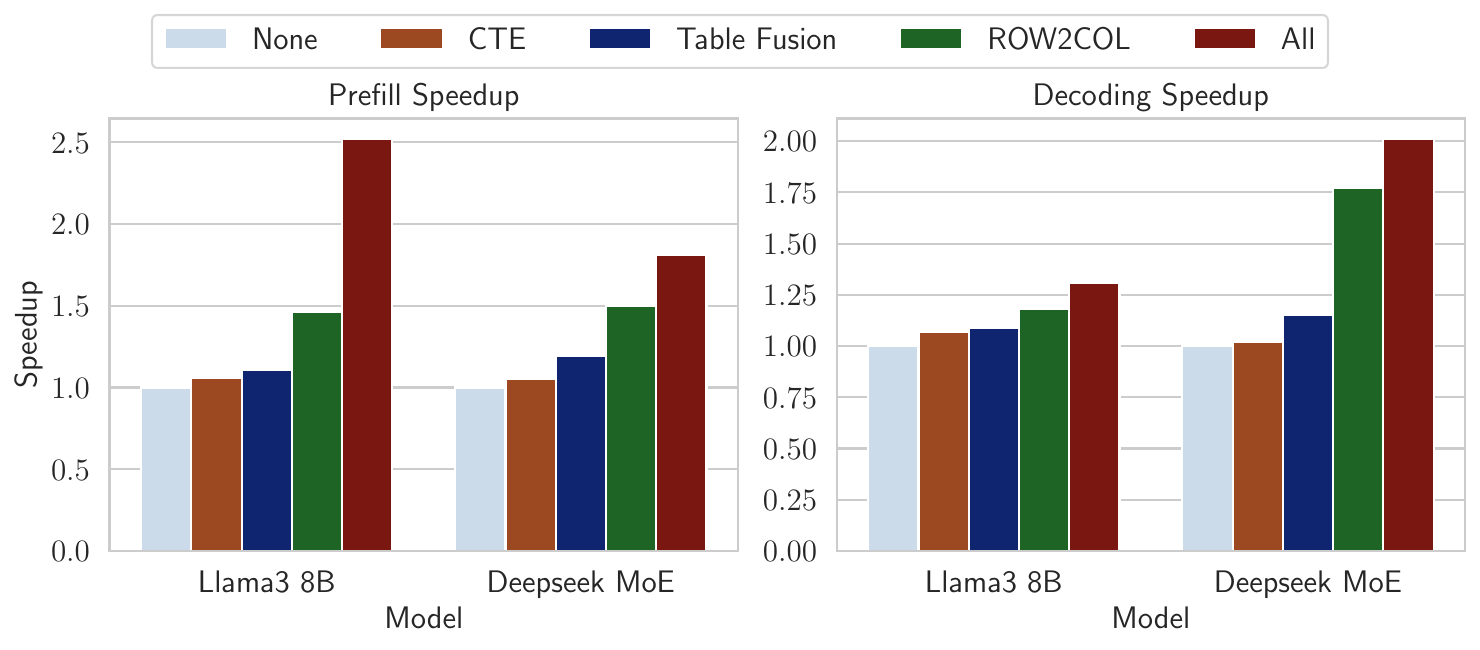}
\caption{ Ablation study for three post optimizations.}
\label{fig:ablation}
\vspace{-5mm}
\end{figure}

\para{Effectiveness of ROW2COL}
In Sec.~\ref{sec:opt}, we introduced \textsc{ROW2COL}, which pivots chunk indices into columns to cut join cardinality and exploit vectorized execution. This creates a trade-off: too many projections in one query can exceed parallelism, while too many subqueries increase I/O. To study this balance, we benchmark speedups over unoptimized SQL across \texttt{\#projection}~$\times$~\texttt{\#subquery} configurations and CPU core counts. The tested matrices are \(25\times4096\) and \(4096\times143{,}336\) with chunk size 64 (64 chunks total).

\begin{figure}[t]
    \centering
    \includegraphics[width=\linewidth]{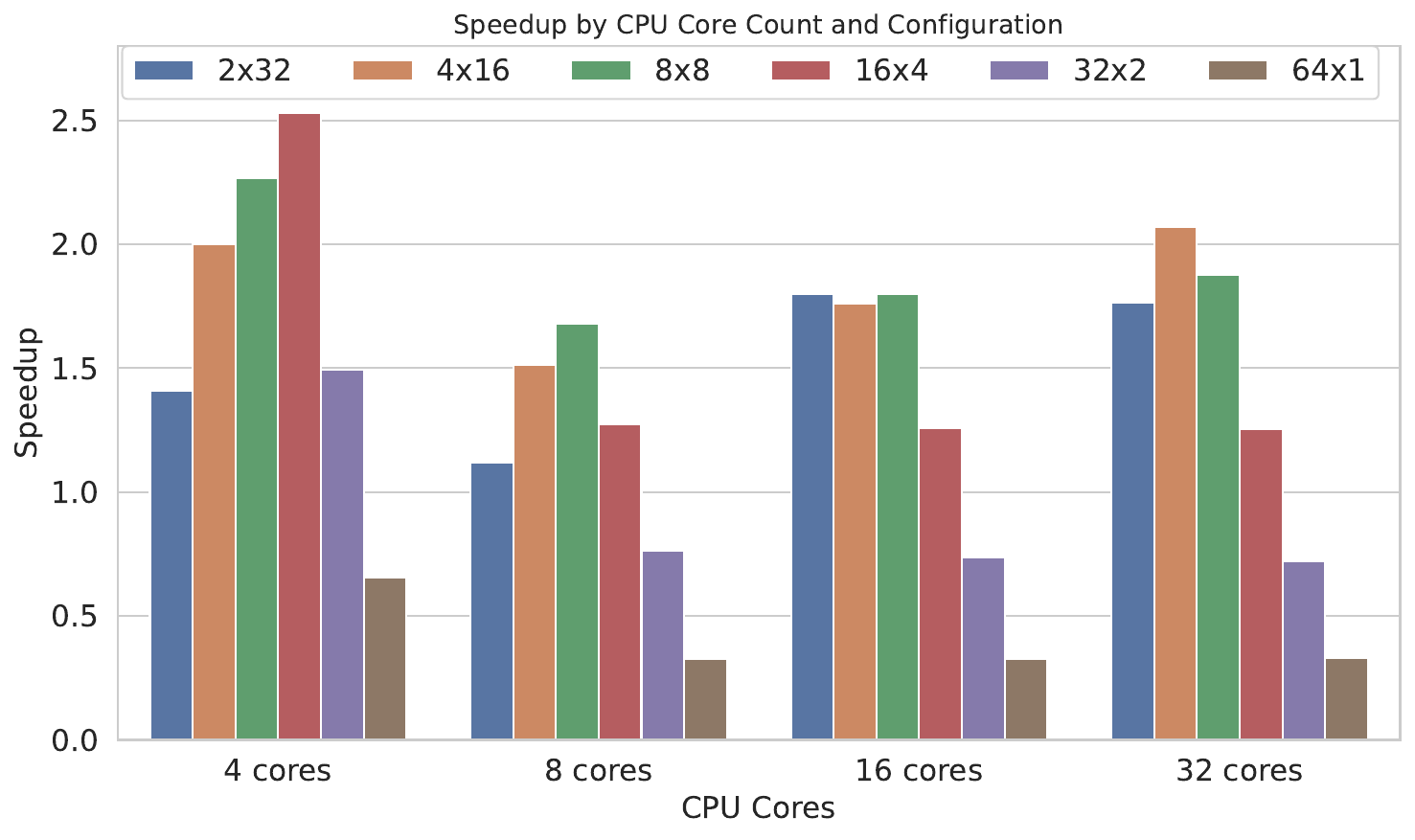}
\caption{\textsc{ROW2COL} optimization improves matrix multiplication performance across most configurations. Based on empirical results for 4 CPU cores, we adopt the 16$\times$4 configuration as a balanced default.}    
\label{fig:row2col}
\vspace{-5mm}
\end{figure}
Fig.~\ref{fig:row2col} shows strong sensitivity to configuration. On 4 cores, 16 projections with 4 subqueries yields the best speedup (2.52$\times$), and most settings still outperform the baseline—confirming \textsc{ROW2COL}'s effectiveness. However, no single configuration is optimal across all cores, and computing all pivoted columns in one query is consistently suboptimal. We therefore fix 16/4 in our experiments and leave a hardware-aware cost model for future work.

\subsection{Influence of Memory Capacity (Q3)}
\label{eval:mem}
 This section analyzes \transql’s memory usage and disk I/O patterns during inference. First, we vary available memory to measure its impact on performance and on the proportion of I/O. Next, we track the total data loaded from disk into memory, revealing the workload imposed by dynamic data paging.

\para{Layer-wise Performance Breakdown} Fig.~\ref{fig:layer_breakdown} breaks down runtime by component (normalization, attention, feed‑forward) for Llama3 and DeepSeek MoE.
\begin{figure}[t]
    \centering
    \includegraphics[width=\linewidth]{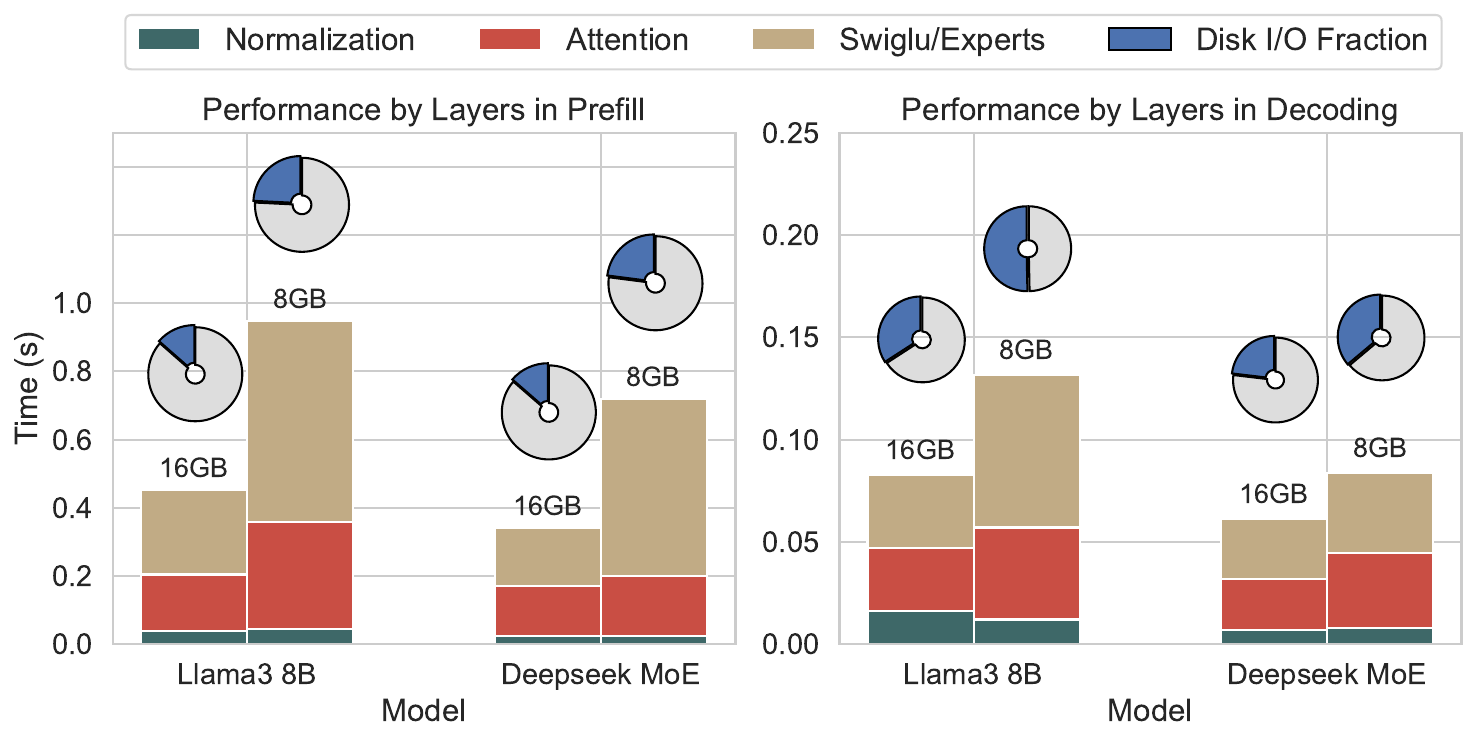}
    \vspace{-2mm}
\caption{ Layer-wise runtime and disk-I/O fraction for 25-token prompts under varying memory. Prefill performance degrades with less memory (nearly 2× disk I/O). MoE decoding scales better since only a subset of experts is active, reducing I/O.
}
\label{fig:layer_breakdown}
\vspace{-4mm}
\end{figure}

In the \textbf{prefill phase}, reducing memory from 16GB to 8GB slows execution, especially in attention and feed‑forward layers.  The pie charts show the disk‑I/O fraction nearly doubling at the lower memory setting. Although I/O is not dominant, each read/write can stall the CPU, so the impact exceeds the fraction alone. DeepSeek MoE exhibits the same pattern, as longer inputs activate more experts, pushing additional weights through the buffer pool and increasing swap traffic. Sufficient memory helps avoid frequent memory–disk swapping and preserves performance.

In the \textbf{decoding phase}, the effect of reduced memory is smaller.  Per‑token work is modest, so latency is far lower than in prefill; however, the same weights must still be streamed, raising the I/O share, which nearly doubles when memory is halved. Absolute latencies for normalization and attention remain stable; the MoE feed‑forward layer is likewise steady because only a small subset of experts is active per token.

%

\para{ Memory Footprint by Layers}  Fig.~\ref{fig:mem_footprint} reports the per‑layer data loaded during prefill; we omit decoding, whose weight I/O closely mirrors prefill. The embedding table is accessed only for prompt tokens, and an index prevents full scans, so its memory footprint is much smaller than the embedding table size. DeepSeek MoE’s expert layer consumes far more memory than Llama3’s SwiGLU because multiple experts are active during prefill. Intermediate activations are also substantial. Under limited memory, these weights and activations are repeatedly swapped between the buffer pool and disk, which drives I/O—56.1GB for DeepSeek MoE and 41.2GB for Llama3 8B. Halving memory roughly doubles the swapped volume and nearly doubles prefill latency (Fig.~\ref{fig:layer_breakdown}). Latency degradation is milder during decoding, since KV vectors are cached and only a single token is processed, making intermediate activations negligible. DuckDB’s buffer management and latency hiding outperform the simple memory‑mapped approach used by our baselines, yielding faster inference with \transql.

\begin{figure}[t]
    \centering
    \includegraphics[width=0.9\linewidth]{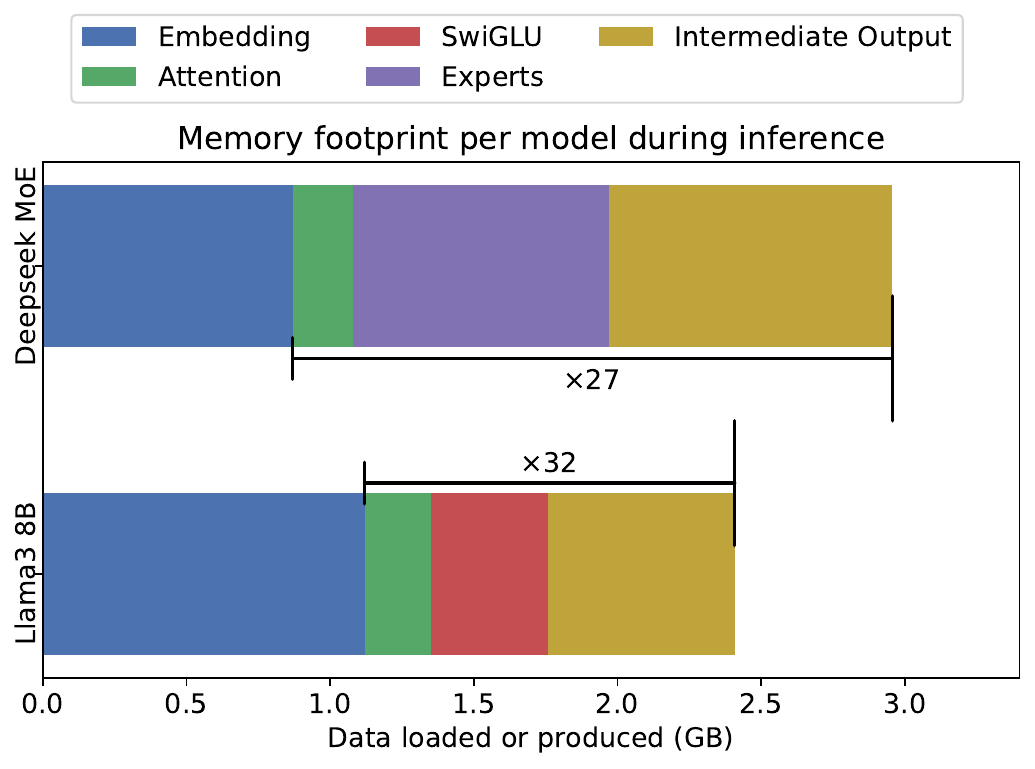}
    \vspace{-2mm}
\caption{
Prefill memory usage with 50-token input.}
\label{fig:mem_footprint}
\vspace{-4mm}
\end{figure}

\vspace{-2mm}
\subsection{Extensibility Beyond Transformers: Convolution as a Case Study (Q4)}
\label{sec:conv}

\begin{figure}[t]
    \centering
    \includegraphics[width=\linewidth]{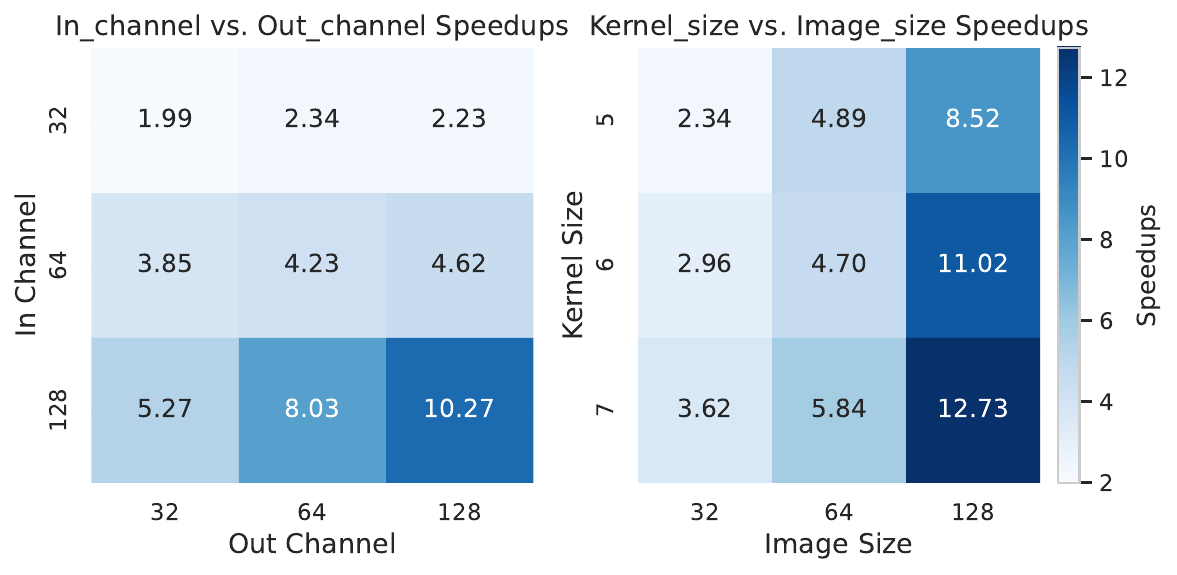}
\caption{In the left figure, kernel size is fixed at 5 and image size at 64; in the right, input channels are fixed at 32 and output channels at 64. SQL generated by \transql consistently outperforms DL2SQL and demonstrates strong scalability.
}    
\label{fig:conv_heatmap}
\vspace{-4mm}
\end{figure}

\transql is designed to be highly extensible to other types of neural network architectures beyond transformers. To demonstrate this generality, we implement a 2D convolution operator within the same template-based SQL generation framework and compare our performance against \textbf{DL2SQL} \cite{dl2sql},  which also translates deep learning computations into SQL. Unlike \transql, DL2SQL does not support self-attention and represents model parameters in a flattened format, lacking the chunk-based representation and vectorization optimizations used in our approach.

Fig.~\ref{fig:conv_heatmap} presents the speedups of  \,\transql with post-optimization over DL2SQL across various convolution configurations. The left heatmap shows speedup trends with increasing input and output channel sizes, while the right illustrates performance changes as image size and kernel size increase. The results clearly indicate that \transql scales more effectively with computational complexity. As both channel dimensions and kernel/image sizes grow, the performance gap widens significantly. For example, when convolving large feature maps (e.g., $128 \times 128$ images with $7 \times 7$ kernels), \transql achieves a speedup of up to \textbf{12.73$\times$} over DL2SQL. This advantage gains from better table design, reduced join cardinality, and improved parallelism through vectorized execution.

Results show that our template-based SQL and post-optimizations extend beyond transformers to convolutions, making \transql viable for a broader range of deep learning inference.

\section{Related Work}
\label{sec:related}
Deep learning has traditionally used specialized hardware and frameworks, but recent work embeds neural computation in relational databases, leveraging database optimizations to express inference in SQL. Here we review in-database inference, memory offloading for large models, and relational approaches to linear algebra, laying the groundwork for our extension to LLMs.

\subsection{In-Database Neural Network Inference}
Early systems \cite{postgreml,sqlserver,verticaML} integrated neural network inference into databases by implementing operations as user-defined functions (UDFs) that offload computation to external ML frameworks. While this decoupled approach avoids reimplementing neural operators in SQL, it incurs substantial cross-system overhead and limits joint optimization of ML inference and SQL processing.

Recent work explores \emph{relation-centric inference}, mapping neural operations to relational query primitives. DuckBrain \cite{duckbrain} and \cite{declaritive} demonstrate SQL-based matrix multiplication for training and inference, suggesting that databases can host neural computation.  \cite{declaritive} targets efficient basic linear algebra on distributed databases with cost estimation and a specialized planner, whereas \cite{duckbrain} exploits a single system’s intrinsic capabilities for both training and inference. Neither supports complex non-linear kernels such as self-attention or residual connections, so we do not compare them directly. Closer to our scope, \textbf{DL2SQL} \cite{dl2sql} expresses each layer as a SQL query over tensor relations, extending beyond basic matrix operations to non-linear kernels (e.g., convolutions, residual connections). Because DL2SQL lacks self-attention, we implement a convolution kernel in \transql\ and compare against DL2SQL’s convolution (Sec.~\ref{sec:conv}). Collectively, these systems underscore a shift toward executing inference inside the database rather than in separate ML frameworks.

Our work builds on this research by extending the relation-centric paradigm to large language models (LLMs). We translate the full suite of LLM operations—including transformer attention and feed-forward layers—into SQL. This approach is, to our knowledge, the first to use a general-purpose SQL engine as the inference engine for LLMs, as earlier in-database systems have focused on smaller models or did not target today’s foundation-scale models.

\subsection{Offloading and Paging Large Models to RAM and Storage}
Running LLMs on commodity or resource-constrained hardware has motivated techniques for memory offloading and paging of model weights. \textbf{FlexGen} \cite{flexgen} enables large models (e.g., 175-billion-parameter GPT-style networks) to run on a single GPU by partitioning data across GPU DRAM, CPU RAM, and disk. \textbf{LLMFlash} \cite{llmflash} leverages high-capacity NVMe flash storage to extend main memory, using techniques like windowing and row--column bundling to reduce data transfers and improve speed. \textbf{FlexInfer} \cite{flexinfer} focuses on on-device inference by dynamically offloading model layers between GPU and CPU, achieving significant throughput improvements on limited-memory devices. All these techniques highlight the benefits of efficient memory management for large models. Our approach adopts a similar strategy but within a database system. Since relational databases have built-in paging mechanisms (e.g., buffer pools and caches), representing LLM weights as relational tables allows the database to transparently manage memory, reducing engineering effort and loosening the coupling with specific hardware architectures.

\subsection{Linear Algebra in Relational Systems}
Our work is also related to research on performing matrix computations with relational query processors. Prior work has leveraged relational databases to execute operations like matrix multiplication \cite{declaritive,chunk_matrix1,chunk_matrix2} and other linear algebra kernels \cite{arun1,arun2} by treating matrices as relational data. Luo \emph{et al.} \cite{luo2018} showed that with minor extensions, a database can serve as a high-performance linear algebra engine, matching the scalability of specialized systems. However, these methods typically focus on blocked-matrix multiplication and struggle with non-linear computations (e.g., softmax, layer normalization) required by LLMs.

In contrast, approaches like tensor relational algebra \cite{TRA,jermaine2021} map tensor operations to relational algebra, enabling some iterative ML tasks to be expressed as SQL queries. Yet, such systems often necessitate substantial changes to underlying compilers and optimizers, limiting their compatibility with the current deep learning ecosystem. Our method, \transql, generates SQL code from an abstract operator level, striking a balance between extensibility, ease of integration, and leveraging existing database engines.

\section{Conclusion and Discussion}
\label{sec:conclusion}

In this work, we present \transql, a template‑based framework that compiles LLM computation graphs into executable SQL for end‑to‑end inference entirely inside a relational database—the first, to our knowledge, to do so without external ML libraries, GPUs, or custom engines. Our results show that principled code generation and optimization enable modern databases to serve as practical engines for resource‑constrained LLM inference. Built purely on standard SQL, \transql is lightweight and portable, easily deployable on laptops, smartphones, and edge hardware, and it integrates smoothly with embedded databases to provide scalable, low‑cost inference without bespoke engineering. While not intended to match GPU frameworks in speed, \transql targets a different trade‑off: accessible, efficient inference on CPU‑only platforms where traditional solutions fall short.

 Although prefill latency remains high—calling for further DB-side optimizations—the fast decoding stage already gives \transql practical value. Recent work offloads only decoding \cite{distserve} or runs a small edge model collaboratively with a large cloud model \cite{collaborative}. Our approach offers an alternative runtime for the edge component: fully portable, SQL-based, and immediately useful in real deployments. Future work includes  enabling quantized arithmetic in databases to further reduce the storage overhead, memory usage, and improve performance. More broadly, relational representations of LLMs may open doors for analytical inspection—treating models as data to better understand their behavior.



\section*{Acknowledgment}
This work has received funding from the Smart Networks and Services Joint Undertaking (SNS JU) under the European Union’s Horizon Europe Research and Innovation programme under Grant No. 101192750. This work is partially supported by NSF award No. 2112631.

\newpage
\bibliographystyle{ACM-Reference-Format}
\bibliography{sample-base}

\newpage

\end{document}